\documentstyle[12pt,aasms4]{article}

\def\arcsecpoint{$''\!.$}
\def\deg{$^{\rm o}$}

\slugcomment{submitted to the {\it The Astrophysical Journal}}

\lefthead{Kraemer \& Crenshaw}
\righthead{Resolved Spectroscopy of the Narrow-Line Region in NGC 1068.
 III, Physical Conditions in the Emission-Line Gas}

\begin{document}

\title{Resolved Spectroscopy of the Narrow-Line Region in NGC 1068.\\
III. Physical Conditions in the Emission-Line Gas\altaffilmark{1}}

\author{Steven B. Kraemer\altaffilmark{2,3},
\& D. Michael Crenshaw\altaffilmark{2,4}}

\altaffiltext{1}{Based on observations made with the NASA/ESA Hubble Space 
Telescope. STScI is operated by the Association of Universities for Research in 
Astronomy, Inc. under the NASA contract NAS5-26555. }

\altaffiltext{2}{Catholic University of America,
NASA's Goddard Space Flight Center, Code 681,
Greenbelt, MD  20771.}

\altaffiltext{3}{Email: stiskraemer@yancey.gsfc.nasa.gov.}

\altaffiltext{4}{Email: crenshaw@buckeye.gsfc.nasa.gov.}

\begin{abstract}

  The physical conditions in the inner narrow 
  line region (NLR) of the Seyfert
  2 galaxy, NGC 1068, are examined using ultraviolet and optical spectra 
  and photoionization models. The spectra were taken with the
  {\it Hubble Space Telescope}/Space Telescope Imaging Spectrograph 
  ({\it HST}/STIS),
  through the 0\arcsecpoint1X52\arcsecpoint0 slit, covering the 
  full STIS 1200 \AA\ to 10000 \AA\ waveband. The slit was centered on a position
  0\arcsecpoint14 north of the optical continuum peak (or ``hot spot'')
  at a position angle (PA) of 202\deg, bisecting the brighter 
  part of the biconical emission-line region. We have measured the 
  emission-line
  fluxes for a region extending 3\arcsecpoint8 northeast ($\sim$ 270 pc) to
  1\arcsecpoint8 southwest ($\sim$ 130 pc) of this point.
  The emission-lines on each side show evidence of two principal kinematic 
  components, one blueshifted
  with respect to the systemic velocity and the other redshifted (the kinematics
  were discussed in a separate paper).
  Based on the photoionization modeling results we find that the physical conditions
  vary among these four quadrants. 1) The emission-line gas in the blueshifted northeast quadrant
  is photoionized by the hidden central source out to
  $\sim$ 100 pc, at which point we find evidence of another source of
  ionizing radiation, which may be due to fast ($\sim$ 1000 km s$^{-1}$)
  shocks resulting from the interaction of the emission-line knots
  and the interstellar medium. Interestingly, this occurs at approximately
  the location where the knots begin to show signs of deceleration. 2) 
  The gas in the 
  redshifted northeast quadrant is photoionized
  by continuum radiation that has been heavily absorbed by gas within $\sim$ 30 pc
  of the central source. We find no strong evidence of the effects of shocks
  in this component. 3) The redshifted emission-line gas in the southwest quadrant is 
  photoionized by unabsorbed continuum from the central source, similar to 
  that in the inner $\sim$ 100 pc of the blueshifted northeast quadrant.
  Finally, 4) the emission-line spectrum of the blueshifted southwest quadrant 
  appears to be
  the superposition of highly ionized, tenuous component within the ionization
  cone and gas outside the cone, the latter photoionized by scattered
  continuum radiation. 
  
  There are several implications of this complicated physical scenario.
  First, the hidden active nucleus is the dominant 
  source of ionizing radiation in the inner NLR. 
  The absorption of continuum radiation along the line-of-sight
  to the redshifted northeast quadrant may result from the intersection
  of the ionization cone and the plane of the host galaxy. 
  Finally, the evidence for
  shock-induced continuum radiation at the point where
  the emission-line knots begin to decelerate indicates that the 
  deceleration is due to the interaction of emission-line knots with 
  slower moving gas, such as the interstellar medium of NGC 1068.

\end{abstract}


\section{Introduction}

 NGC 1068 is the nearest ($z =$ 0.0038) and brightest Seyfert 2 galaxy and has 
 been observed extensively in all wavebands from the radio to the X-ray
 (it was recently the subject of an entire workshop, see Gallimore 
 \& Tacconi 1997). The detection of broad (Full Width Half Maximum, 
 {\it FWHM} $\approx$ 4500 km s$^{-1}$) permitted emission lines in the polarized
 flux from NGC 1068 (Antonucci \& Miller 1985; Miller, Goodrich, \& Mathews
 1991) was the inspiration for the unified model for Seyfert galaxies,
 the basis of which is that all Seyfert 2s harbor a broad-line region (BLR) 
 and non-stellar continuum source, which are hidden along our line-of-sight by a dense molecular torus, while
 the central source and BLR are viewed directly in Seyfert 1s
 (see Antonucci 1993).

 Ground-based narrow-band images of the narrow-line region (NLR) of NGC 1068
 reveal that the emission lines form a biconical structure, roughly
 parallel to the radio axis, extending northeast and 
 southwest of the nucleus (Pogge 1988). Observations with the 
 {\it Hubble Space Telescope} (HST) reveal that the inner NLR contains
 numerous knots and filaments of ionized gas, also in a biconical
 geometry (Evans et al. 1991; Macchetto et al. 1994). 
 The hidden active galactic nucleus (AGN) is thought to lie 0\arcsecpoint13 
 south of the optical continuum 
 peak (Capetti, Macchetto, \& Lattanzi 1997), and is thought to be
 associated with the thermal radio source designated S1 (cf.
 Gallimore et al. 1996).

 Analysis of ground-based observations has revealed that the emission-line
 gas in the extended NLR, $\sim$ 1 kpc from the
 nucleus, is most likely photoionized by continuum radiation
 emitted by the hidden AGN (Evans \& Dopita 1986). Although it
 follows that photoionization must be important, if not dominant,
 closer to the active nucleus, conditions 
 within $\sim$ 200 pc of the nucleus are apparently quite complex. Inspired
 by the surprisingly strong C~III $\lambda$977 and N~III $\lambda$990 lines
 seen in {\it Hopkins Ultraviolet Telescope} spectra, Kriss et al. (1992)
 argued that shock heating is an important process in the inner NLR
 (but see Ferguson, Ferland, \& Pradhan [1994] for another interpretation).
 Based on HST/Faint Object Camera (FOC) long slit spectra, Axon et al. (1998) 
 found evidence for an increase in emission-line flux and ionization state at 
 a region $\geq$ 150 pc northeast of the nucleus, which may be 
 due to the interaction of the radio jet which traverses the NLR of NGC 1068 
 (Wilson \& Ulvestad 1983)
 with interstellar gas clouds. In fact several authors have argued that
 the ionizing radiation emitted by fast ($\sim$ 500 km s$^{-1}$) shocks, 
 arising either from cloud/cloud interactions (cf. Sutherland, Bicknell,
 \& Dopita 1993), or jet/cloud interactions (Wilson
 \& Raymond 1999; Morse, Raymond, \& Wilson 1996), may power the 
 NLR in Seyfert 2 galaxies. On the other hand, using
 HST/Faint Object Spectrograph data, Kraemer, Ruiz \& Crenshaw
 (1998) examined conditions in the inner $\sim$150 pc of the NLR and 
 determined that the 
 observed emission lines could generally arise in gas photoionized by the 
 hidden AGN (although there appeared to be some additional collisional
 excitation in the region nearest the radio jet). Recently,
 Alexander et al. (2000) found evidence that the ionizing radiation
 incident upon the NLR was heavily absorbed by an intervening layer of gas 
 (although it should be noted that the spectra these authors analyzed
 were from an area encompassing nearly the entire NLR and, thus,
 any local effects were diluted). In summary, although it is likely
 that the NLR is photoionized, both the spectral
 energy distribution (SED) of the ionizing radiation incident
 on the emission-line clouds and the physical process(es) in 
 which this radiation arises have yet to be firmly established.
 
 In previous papers, we used STIS long slit spectra, taken
 along a position angle of 202\deg, to examine the nature of the
 extended continuum radiation in the NLR (Crenshaw \& Kraemer 2000a, hereafter
 Paper I), the physical conditions near the continuum hot spot
 (Kraemer \& Crenshaw 2000, hereafter Paper II), and the NLR kinematics
 (Crenshaw \& Kraemer 2000b, hereafter CK2000). Interestingly, we
 found that the emission-line gas near the hot spot, in spite of the
 presence of coronal lines
 such as [Fe~XIV] $\lambda$5303 and [S~XII] $\lambda$7611, appears to be
 photoionized by the central continuum source, with no evidence of 
 additional shock heating (Paper II).  In this paper, we use the same long slit 
 dataset and 
 photoionization models to determine the physical conditions as a
 function of position within the
 inner $\sim$ 200 pc of NLR of NGC 1068. In Section 3 we will discuss
 general trends derived from the emission-line flux ratios, specifically
 reddening and ionization state at different radial positions. In section
 4 and 5 we will present the details of photoionization models of the
 emission-line gas, which reveal that the physical conditions
 vary greatly along different sight-lines to the active nucleus. Finally, 
 we will discuss the implications of these results, including
 a possible connection between the NLR conditions and the orientation
 of the emission-line bicone and the host galaxy. We adopt a
 systemic redshift of cz $=$ 1148 km s$^{-1}$ from H~I observations
 (Brinks et al. 1997) and a distance of 14.4 Mpc (Bland-Hawthorn et al. 1997), so
 that 0\arcsecpoint1 corresponds to 7.2 pc.

\section{Observations and Analysis}

We observed NGC 1068 with {\it HST}/STIS using a 52$''$ x 0\arcsecpoint1 slit to 
obtain spectra over 1150 -- 10,270 \AA\ at a resolving power of 
$\lambda$/$\Delta\lambda$ $\approx$ 1000. 
Our slit position (at PA $=$ 202\deg, offset 0\arcsecpoint14 north of the
continuum peak) intersects a number of bright 
emission-line knots in the inner NLR (see Paper I).
We extracted spectra using a bin length of 0\arcsecpoint2 (14.4 pc) along the slit, to 
obtain reasonable signal-to-noise ratios for the emission lines.
Paper I gives additional details on the observations and data reduction. 

In CK2000, we showed that the [O~III] $\lambda$5007 
emission splits into two major kinematic components (blueshifted and redshifted 
relative to the systemic velocity) both NE and SW of the optical continuum 
peak (hot spot). The 
other emission lines show the same general behavior (see Paper I). Since the 
[O~III] $\lambda$5007 emission is isolated from other lines and has the greatest 
signal-to-noise over this extended region, we used it as a template for the 
other lines. Thus, at each spatial bin, we fit the [O III] 
emission with one or two Gaussians, and determined the velocity positions and 
widths of each component. These positions and widths were then kept fixed for 
the other emission lines, and the height of the Gaussians were allowed to vary 
independently to determine the fluxes. The advantage of this procedure is that 
it allows us to determine a consistent set of line ratios for different 
kinematic components along the same line of sight.
We can therefore place the emission-line measurements into one of four 
quadrants: NE-blue, NE-red, SW-blue, SW-red. The location of the emission is 
further specified by the projected distance (in arcseconds) from the center of 
the bin that contains the hot spot.

Our fitting procedure works well for lines that are relatively isolated (e.g., 
H$\beta$) or for doublets with fixed ratios (e.g., C~IV $\lambda\lambda$1548.2, 
1550.8). However, many other features, such as the blend of H$\alpha$ and N~II 
$\lambda\lambda$6548, 6584 or the blend of H$\gamma$ and [O~III] $\lambda$4363, 
are impossible to disentangle in this fashion, due to the mixing of kinematic 
components of different lines. Therefore, we only measured isolated lines that 
could yield reliable fluxes for each kinematic component. Errors in the fluxes 
are from the uncertainties in the Gaussian fits and different reasonable 
continuum placements (added in quadrature). As in the past, we used the He~II 
$\lambda$1640/$\lambda$4686 ratio to determine the reddening (see Paper II), and 
dereddened the observed ratios using a standard extinction curve (Savage and 
Mathis 1979). Errors in the dereddened ratios include a contribution from the 
reddening error (also added in quadrature). The observed line ratios are 
listed in Table 1a, while the dereddened
line ratios, H$\beta$ fluxes, and reddening (E$_{B-V}$) are listed in 
Tables 1b. Note that our central bin, position 0\arcsecpoint0, is
approximately 0\arcsecpoint3 north of S1.

\section{General Trends}

\subsection{Reddening}
 
 Figure 1a shows E$_{B-V}$ as a function of position along the 
 slit. Although the reddening shows considerable variation within
 each quadrant, it is apparent that the emission-line gas in the NE-blue 
 quadrant, with an average E$_{B-V}$ is $\approx$ 0.35, is more
 heavily reddened than that in the NE-red quadrant, where the 
 average E$_{B-V}$ is $\approx$ 0.22 (there are not enough
 datapoints in the SW quadrants for us to determine trends).
 It is possible that both regions are viewed through
 an external screen of dust, corresponding to a reddening of
 E$_{B-V}$ $=$ 0.2 and a column of
 hydrogen of $\sim$ 10$^{21}$ cm$^{2}$, assuming a typical
 Galactic gas-to-dust ratio (cf. Shull \& Van Steenberg 1985). 
 The NE-blue quadrant must be viewed through an additional column of
 $\sim$ 7 x 10$^{20}$ cm$^{-2}$.
 The simple kinematic model described in CK2000 requires that
 the NE-blue quadrant lies in front of the NE-red quadrant along our
 line-of-sight. Such a geometry can fit the observed reddening
 if the difference in the reddening
 is associated with the individual components and the total covering 
 factor of the NE-blue components is
 small enough that it does not eclipse much of the NE-red quadrant.

 As we will demonstrate in the
 following sections, the emission-line gas in the NE-blue
 quadrant is generally matter-bounded (optically thin). Therefore,
 we are able to constrain the column density of the emission-line clouds,
 and hence their spatial extent across our line-of-sight. The extent
 of the emission-line gas is quite small compared to the size of the spatial 
 bins ($<$ a few \%); hence it is likely that we are not viewing the NE-red 
 components through a screen of material in the NE-blue quadrant. 

 The addditional reddening might be associated with the
 emission-line gas itself if we are preferentially viewing the
 the ionized face of the NE-red components while viewing the
 backends of the NE-blue components, which is consistent with our
 proposed geometry in CK2000 (the
 NLR gas in NGC 4151 is viewed in a similar fashion [Kraemer et al. 2000]).
 However, as we will show, there is insufficient dust embedded in the 
 emission-line 
 clouds to fully account for this. Another possibility is that
 the clouds are filamentary rather than spherical, in which case
 the path-length for photon escape may be quite large. However, this
 requires that the NE-red components are either more symmetric or, again,
 that we are preferentially viewing their illuminated surfaces.
 Note that, in the NE-red components, the average dereddened L$\alpha$/H$\beta$ is
 close to Case B (24 -- 36; Osterbrock 1989), 
 as shown in Figure 1b, which indicates minimal destruction of L$\alpha$
 by dust within the emission-line gas (although there are several
 cases where the suppression is significant, as noted Section 5.3).
 Unfortunately, we were unable to separate the intrinsic and geocoronal
 L$\alpha$ in the NE-blue components.
 
 In summary, while this difference in 
 reddening does not compromise our kinematic model, we cannot
 determine its source.

\subsection{Density and Ionization}

 Although we will use photoionization models in order to make
 a detailed examination of the physical conditions of several components
 of the emission-line gas (see next section), we can infer
 general trends from the [O~III] $\lambda$5007/H$\beta$
 and [O~II] $\lambda$3727/H$\beta$ ratios as a function of 
 position along the slit. These ratios are shown in Figures 1c and 1d.
 
 In the NE-blue quadrant, [O~III]$\lambda$5007/H$\beta$ increases slightly
 between $-$0\arcsecpoint2 and $-$0\arcsecpoint8, which is followed by a slight
 drop out to $-$1\arcsecpoint2, then a steeper increase out to $-$2\arcsecpoint0.
 The [O~II] $\lambda$3727/H$\beta$ ratio increases steeply from $-$0\arcsecpoint2
 to $-$1\arcsecpoint2, followed by an abrupt decrease. One possible
 interpretation is that the density is high enough (well above the
 critical density of  3.3 x 10$^{3}$ cm$^{-3}$;
 Osterbrock 1989) that [O~II] $\lambda$3727 is significantly suppressed
 by collisional de-excitation in the inner 0\arcsecpoint6. Apparently, the density
 decreases with the radial distance, as D$^{-2}$,
 since 
 [O~III] $\lambda$5007/H$\beta$ is fairly constant, out to $-$1\arcsecpoint4,
 with [O~II] $\lambda$3727/H$\beta$ increasing as the density drops.
 At this point, there is a simultaneous rise in [O~III] $\lambda$5007 and drop
 in [O~II] $\lambda$3727, which appears to be the result of an increase in the
 ionization state of the gas (the relative strength of these lines
 is a good indicator of the ionization state in low density gas 
 (Ferland \& Netzer 1983), which could be the result of a faster
 density drop-off or an increase in the flux of ionizing radiation; we prefer
 the latter, as we will show in Section 5.2. 

 In the NE-red quadrant, [O~III] $\lambda$5007/H$\beta$ is significantly
 lower than in the NE-blue quadrant, which indicates that
 this gas is in a lower state of ionization, a point that
 we will discuss in Section 5.3. [O~III] $\lambda$5007/H$\beta$
 decreases slightly out to $-$1\arcsecpoint6 (with one exception), then recovers 
 somewhat, while [O~II] $\lambda$3727/H$\beta$ shows the same overall trend as 
 in the NE-blue quadrant. Hence, density appears to be decreasing with radial
 distance. Unlike in the NE-blue quadrant, there is no abrupt increase
 in relative strength of [O~III] $\lambda$5007, hence there is no 
 strong evidence of an increase in the ionization state of the gas
 between $-$1\arcsecpoint4 and $-$2\arcsecpoint0. The large values of 
 [O~III]$\lambda$5007/H$\beta$ at radial distances beyond $-$3\arcsecpoint0 
 is most likely indicative of
 the presence of large amount of optically thin gas, as seen in the 
 outer NLR of NGC 4151 (Kraemer et al. 2000). 

 There are too few datapoints in the SW quadrants to clearly determine trends,
 due to the overall weakness of emission in this direction. A possible
 explanation, consistent with our kinematic model (CK2000), is that
 the majority of the SW cone lies behind the galactic plane.
 The weakness or absence of [O~II] $\lambda$3727 in the SW-red
 quadrant, although possibly just a detectability problem, is also 
 consistent with higher densities than in either of the NE quandrants.

\section{Photoionization Models}

   The details of our photoionization code have been discussed in
   several previous publications (Kraemer 1985; Kraemer \& Harrington 1986;
   Kraemer et al 1994). 
   As usual, the photoionization models are parameterized in terms of the 
   dimensionless ionization parameter, U, which is the number of ionizing photons
   per hydrogen atom at the illuminated face of the cloud. The inputs to
   the models include the gas density (n$_{H}$), the distance from the
   nucleus (D), the number of ionizing 
   photons (Q), the spectral energy distribution of the ionizing
   radiation, the elemental abundances, the dust/gas ratio, and the
   column density of the emission-line clouds. 

   The possibility
   of non-solar elemental abundances in the inner NLR of NGC 1068
   has been discussed previously (Netzer 1997; Netzer \& Turner 1997;
   Kraemer et al. 1998). In Paper II, we argued that there
   was no strong evidence for non-solar O, Ne, and N, and that, although
   the Fe/O ratio did appear to be supersolar within 30 pc of the
   hidden nucleus, this might not be true at larger radial distances. 
   Therefore, we have chosen to assume solar 
   abundances (cf. Grevesse \& Anders 1989) for these models. The numerical 
   abundances, relative to hydrogen, are as follows: He$=$0.1,
   C$=$3.4x10$^{-4}$, O$=$6.8x10$^{-4}$, N$=$1.2x10$^{-4}$, Ne$=$1.1x10$^{-4}$, 
   S$=$1.5x10$^{-5}$, Si$=$3.1x10$^{-5}$, Mg$=$3.3x10$^{-5}$, Fe$=$4.0x10$^{-5}$.
   
   We have assumed that both silicate and carbon dust grains are present,
   although the dust-to-gas ratio, which we express as a dust ``fraction''
   relative to that in the
   interstellar medium (ISM) of the Milky Way (cf. Draine \& Lee 1984), varies 
   throughout the emission-line region. The depletions of elements from gas phase are equal to the
   dust fraction multiplied by the following ISM depletions (cf. Snow \& 
   Witt 1996): C, 65\%; O, 50\%; Si, Mg, and Fe, 100\%.

   Although Alexander et al. (2000) report
   evidence of the presence of a Big Blue Bump in the SED of the ionizing 
   continuum, in Paper II we demonstrated that the emission-line
   spectrum from the hot spot could be produced by a simple
   power-law continuum.
   Here we assume the same ionizing continuum used in Paper II, specifically
   a broken power-law of the form, F$_{\nu}$ $=$ K$\nu^{-\alpha}$, 
   as follows (also, see Figure 3):

\begin{equation}
    \alpha = 1.0,~ h\nu < 13.6~eV
\end{equation}
\begin{equation}
    \alpha = 1.4,~ 13.6~eV \leq h\nu < 1000~eV
\end{equation}
\begin{equation}
    \alpha = 0.5, ~h\nu \geq 1000~eV
\end{equation}
   The intrinsic luminosity above the
   Lyman limit, $\sim$ 3 x 10$^{44}$ erg s$^{-1}$ (Q $\sim$ 4x10$^{54}$ photons 
   s$^{-1}$, is typical of Seyfert 1 nuclei (Pier et al. 1994).
   As we will discuss in the following sections, there is
   evidence at specific locations that the ionizing continuum
   has been modified by either 1) absorption by intervening gas, or
   2) an additional component of UV -- X-ray radiation, such as that produced by a
   fast shock.    

   For each position, we found the best fit was obtained by a two-component
   model, with the density (n$_{H}$) and the dust fraction of each left as free
   parameters. Multiple component models were used with success in our earlier
   study of the NLR in NGC 1068 (Kraemer et al. 1998), while
   Kraemer et al. (2000) and Schulz \& Komossa (1993) 
   found strong evidence for components of different densities at the same radial distances
   to model the NLR and extended NLR, respectively, in NGC 4151 (one
   might infer that local density inhomogeneity is a typical characteristic
   of the NLR). To determine the distance between the emission-line clouds and 
   the central source, D, we measured the angular separation of the center of 
   each extraction bin and the location of the S1 radio source (see
   Paper II, and references therein). We did not correct
   for projection effects, which would only increase D by a factor
   $\leq$ 1.4 (CK2000). We chose the dust fraction to account for 
   the suppression of resonance lines (e.g., L$\alpha$ and
   C~IV $\lambda$), but were constrained by the strength of emission lines such as Mg~II
   $\lambda$2800 and [Fe~VII] $\lambda$6087, whose presence
   requires a portion of these elements to be in gas phase.

   Similar to the approach taken in Kraemer et al. (1998),
   we have assumed
   that one component is screened from the ionizing source by the other,
   the denser gas by the 
   the more tenuous component. In such a model, the covering factor of the
   dense component with respect to the continuum source is 
   constrained by the covering factor of the tenuous gas by which it is
   screened. Note that this is different than the
   geometry assumed in Paper II, in which the low- and high-ionization gas
   were not co-planar.
   Such a geometry is 
   plausible in the case of the hot spot, since the low ionization lines were 
   redshifted with the respect to the high ionization lines. However, in the 
   extended emission-line gas, there are low and high ionization lines
   with the same velocities, which implies that the regions of 
   different ionization are kinematically associated. The total 
   column density, N$_{H}$ (the sum of the columns of ionized and neutral 
   hydrogen), of the tenuous
   component was fixed to provide the best fit to the He~II lines, while
   that of the denser component was constrained by the observed 
   strength of the [O~I] $\lambda$6300 line.

   Our photoionization code does not include the pumping
   of UV resonance lines by scattering of continuum radiation and 
   continuum fluorescence (cf. Ferguson, Ferland, \& Pradhan 1994). 
   In modeling the hot spot spectrum, we generated
   comparison models with CLOUDY90 (Ferland et al. 1998), and did not find 
   this effect important for C~IV $\lambda$1550, at turbulent velocities
   $<$ 500 km s$^{-1}$. The reasons are twofold. First, for very 
   optically thick lines ($\tau$$_{line center}$ $>$ 100) the
   pumping efficiency is low. Second, in typical NLR conditions 
   (e.g., a cloud with n$_{H}$ $=$
   10$^{4}$, U $=$ 10$^{-2}$, which is optically thick at the He~II Lyman
   limit), the contribution to 
   C~IV $\lambda$1550 from collisional excitation dominates any
   boost from resonance scattering. However, in Paper II, we find that resonance scattering
   is an important effect for other UV resonance lines, such as 
   C~II $\lambda$1335 and N~V $\lambda$1240, for which collisional
   excitation is less dominant, due to weak collision strengths or smaller 
   ionic columns. Hence, the predicted N~V $\lambda$1240 flux
   is a lower limit, since pumping can boost the relative strength of this
   line by a factor of 2 -- 3 for the conditions in typical 
   emission-line gas and turbulent velocities $<$ 500 km s$^{-1}$ (we
   do not expect that the turbulent velocities are even this high,
   given the fact that the {\it FWHM} of the emission lines
   is $<$ 1000 km s$^{-1}$ and the broadening is probably partially due to the 
   superposition of different kinematic components).

\section{Model Results}

\subsection{NE-blueshifted Quadrant -- Photoionization by the AGN}

  As noted above, we initially assumed that the blueshifted emission-line 
  gas NE of the optical continuum peak is ionized solely by the
  continuum radiation emitted by the hidden AGN. To test
  this, we first fit the emission-line spectrum from the bin centered
  at $-$0\arcsecpoint2. The physical parameters for
  the two components, and a comparison of the model prediction to the
  observed line ratios are given in Table 2\footnote[5]{In Table 2 -- 14, we 
  list, in addition to the model parameters described in Section 4, the following:
  the predicted emitted H$\beta$ flux (F$_{H\beta}$); the emitting area,
  which is the observed H$\beta$ luminosity divided by F$_{H\beta}$; the
  Depth, which is a lower limit to the distance the emitting region projects into the plane
  of the sky, constrained by the slit width (7.2 pc); and the 
  fraction each component contributes to the total H$\beta$ flux.}.
  Assuming equal contributions to the H$\beta$ flux
  from each component, we obtained a very good fit for each
  emission line (the model fits the
  data within a factor 2 and, generally, much better than that). The
  one exception is N~V $\lambda$1240, which is often poorly fit,
  even with the expected boost from resonance scattering. We have
  discussed the underprediction of this line by photoionization models 
  (Kraemer et al. 2000; Paper II), although we have no ready explanation. 

  The line ratios do not vary 
  significantly among the bins from $-$0\arcsecpoint2 to $-$1\arcsecpoint2
  (see Table 1b),
  which implies that the density is falling off roughly as
  D$^{-2}$. To test this, we generated a model of the
  emission-line spectrum in the Bin at $-$1\arcsecpoint0, lowering
  the densities of the two components accordingly (we increased N$_{H}$
  by a factor of 3 to match the strength of the [O~I] $\lambda$6300 line).
  The comparison of the model and the observed line ratios is given in
  Table 3. Again the fit is reasonably good. The overprediction of 
  C~IV $\lambda$ 1550 and Mg~II $\lambda$2800 could
  be easily rectified by assuming a slightly higher dust fraction than the
  10\% assumed here (hence, we assumed a somewhat higher dust fraction
  for the emission-line gas further out). Although the N~V $\lambda$1240
  might be fit with a turbulent velocity of 500 km s$^{-1}$, this would
  worsen the C~IV $\lambda$1550 fit.

\subsection{NE-blueshifted Quadrant --  AGN $+$ Shock}

  Although irradiation by the central source
  appears to be the dominant process in the inner 120 pc, it does not seem
  to fully explain the emission-line spectrum near $-$1\arcsecpoint4, which
  show a large He~II $\lambda$4686/H$\beta$ ratio and strong
  [Ne~IV] $\lambda$2423 and [Ne~V] $\lambda$3426 lines, as shown in
  Figures 2a -- c. 
  The increase in the He~II $\lambda$4686/H$\beta$ ratio could be
  explained by a greater fraction of optically thin, tenuous gas, however
  such conditions would result in weak [Ne~IV] $\lambda$2423 and, perhaps,
  [Ne~V] $\lambda$3426.
  Also, the relative strengths of the low ionization lines, such as 
  [O~I] $\lambda$6300 and [O~II] $\lambda$3727, are not appreciably different 
  than
  in the emission-line gas closer to the nucleus, which constrains the
  fraction of more highly ionized gas.

  As noted in Section 1, Axon et al. (1998) discussed the possibility of an additional source of 
  ionizing radiation in the inner NLR of NGC 1068, perhaps arising from the 
  interaction of the emission-line clouds or radio jet
  with the interstellar medium. Models of fast shocks (V$_{shock}$ $\geq$ 400 
  km s$^{-1}$) predict strong EUV and X-ray continuum and line emission from 
  the shock front (cf. Sutherland et al. 1993; Morse et al. 1996).
  Since the emission-line clouds at radial distances
  $>$ 100 pc appear to be decelerating from maximum velocities of
  $\sim$ 1000 km s$^{-1}$ (given projection effects; CK2000), it is entirely 
  plausible that such fast shocks arise. Recent STIS longslit
  observations, which mapped the NLR of NGC 1068 at a resolving
  power of $\lambda$/$\Delta$$\lambda$ $\approx$ 10,000, show
  local velocity structure which may be evidence of cloud disruption
  due to shocks (see Cecil et al. 2000).

  The combined effects of shocks and photoionization by
  an AGN have been modeled by Viegas-Aldrovandi \& Contini (1989a and b, and 
  references therein to their earlier work). However, these 
  were generated for a specific range size and structure of the emission-line
  clouds, and are not readily applicable to the present data. Although
  our photoionization code does not include shocks, we
  have approximated its effect by combining the continuum 
  from the hidden AGN, 
  appropriately diluted by
  distance, with the ionizing radiation from the shock front.
  This was done in a simplistic fashion, by assuming that
  both the AGN radiation and shock-induced radiation are incident
  upon the same ``face'' of the cloud. Although this is almost certainly
  not the case for radially outflowing clouds, it is sufficient to
  demonstrate the effects of an additional source of UV -- X-ray radiation. 
  For the shock-induced radiation, we used models generated by
  Wilson \& Raymond (1999) for shock velocity of 1000 km s$^{-1}$
  (see Figure 4). The observed 2 -- 10 keV flux is $\sim$ 2 -- 8
  x 10$^{-12}$ ergs cm$^{-2}$ s$^{-1}$, or 0.5 -- 2.0 x 10$^{41}$ ergs s$^{-1}$,
  assuming a distance of 14.4 Mpc (Bland-Hawthorn et al. 1997), which is 
  consistent with the X-ray emission
  produced by the starburst disk in NGC 1068 (Wilson et al. 1992; Turner et al. 
  1997). Since the starburst is likely to contribute most of the
  extended X-ray flux, we scaled to a total shock
  luminosity above the Lyman limit to $\sim$ 10$^{42}$ ergs cm$^{-2}$ 
  sec$^{-1}$, ($\sim$ 5 x 10$^{40}$ ergs s$^{-1}$ in the 2 -- 10 keV band), 
  which can be produced by an emitting area of $\sim$ 10$^{39}$ cm$^{2}$ for
  a precursor gas density of n$_{H}$
  $=$ 10$^{3}$ cm$^{-3}$.
  Of course, the shock-induced luminosity could be much lower.
       
  For the model of the $-$1\arcsecpoint4 bin, 
  we assumed that the shock front contributes $\sim$2.5 times the ionizing
  flux of the central source, which, given our scaling of the shock,
  requires distance from the shock of 10 pc (i.e., within an extraction
  bin). The comparison of the model and observations is given in Table 4. 
  For the bin centered at $-$1\arcsecpoint8, we
  assumed the shock contributes $\sim$ 2.8 times the flux of the central
  source. The results are given in Table 5. 
  Considering the crude manner in which we included the effects of the
  shock, the model predictions fit the observed emission-line ratios
  reasonably well. The underpredictions of [Fe~VII] $\lambda$6087 may 
  indicate that less iron is depleted onto dust grains than we 
  assumed. The underprediction of the [Ne~IV] $\lambda$2423 is slightly
  troubling, although the fit is only marginally outside our limit
  of acceptability, since its relative strength was one of the spectral
  features that suggested a boost in ionization.
  The shock induced emission is primarily a combination of
  free-free continuum and recombination lines (Sutherland et al. 1993),
  and the frequency grid in our photoioization models is somewhat coarse
  (Kraemer 1985); hence, we may have missed important emission 
  features above 63.5 eV (the ionization potential for Ne~III). Alternatively,
  the strength
  of collisionally excited UV lines can be enhanced in high temperature
  (2 x 10$^{4}$ K -- 10$^{5}$ K) zones behind the shock front (cf.
  Allen, Dopita, \& Tsevtanov 1998), an effect which we have not included,
  although one would expect that would result in underpredictions of 
  C~IV $\lambda$1550 and [Ne~V] $\lambda$3426. Of course, the fit for
  [Ne~IV] $\lambda$2324 would have been even worse without 
  including the shock-induced ionizing radiation.
  
\subsection{NE-redshifted Quadrant}
  
 Based on the observed emission-line ratios (e.g.,  
 [O~III] $\lambda$5007/H$\beta, $He~II $\lambda$4686/H$\beta$, and 
 [O~III] $\lambda$5007/[O~II]$\lambda$3727; Figures 1c, 2a, and 2d,
 respectively), it is
 apparent the NE-red gas is generally
 in a lower state of ionization than the NE-blue gas.
 There are several ways in which this might occur. One is if the redshifted
 gas is actually at larger radial distance from the central source than the
 corresponding blueshifted gas. However, the 
 kinematics of the [O~III] emission-line gas suggests that the bicone
 axis is nearly in the plane of the sky (CK2000), hence the red- and 
 blueshifted components along any sight-line are probably at roughly the
 same distance from the nuclear source. Although the lower ionization
 state of the redshifted gas could be due to its having a higher
 density than its blueshifted counterpart, the presence of 
 [O~II] $\lambda$3727 in
 both components would suggest otherwise (although we do find that
 the redshifted components may be somewhat denser). A more plausible explanation
 is that the redshifted gas is exposed to a lower flux of ionizing radiation.
 Although this could be the result of an intrinsically anisotropic radiation
 field, we think it is more likely the result of a
 component of gas close ($<$ 30 pc) from the central source, which
 absorbs some fraction of the ionizing radiation emitted towards
 the redshifted gas. Notably, if there is mix of optically thin and thick
 components among the emission-line gas, as is the case in the NE-blue
 quadrant, the lower value of the He~II $\lambda$/H$\beta$ ratio is
 evidence for a paucity of photons above the He~II Lyman limit, which
 can be caused by an intervening absorber (Alexander et al. 1999;
 Kraemer et al. 2000).

 As mentioned in Section 1, Alexander et al. (2000) found strong evidence that 
 the narrow-line gas
 in NGC 1068 is irradiated by continuum which has been partially absorbed by 
 a neutral hydrogen column of
 $\sim$ 6 x 10$^{19}$ cm$^{-2}$. Similar to the situation for NGC 4151 (
 Kraemer et al. 2000),
 we modeled the 
 absorber as a large column of highly ionized
 gas (an ``X-ray'' absorber) and a thinner, outer layer of low ionization
 gas (a ``UV'' absorber). 
 For the X-ray absorber, we assumed U $=$ 1.0 and N$_{H}$ $=$ 3.7 x 10$^{22}$ 
 cm$^{-2}$, while for the UV absorber, U $=$ 10$^{-3.7}$ and N$_{H}$ $=$
 1.0 x 10$^{19}$ cm$^{-2}$. The effects of the absorbers on the ionizing
 continuum are shown in Figure 3, with deep edges due to H~I, at 13.6 eV, and 
 He~II, at 54.4 eV, primarily from the UV absorber, and the combined effects 
 of O~VII and O~VIII, above 740 eV, from the X-ray absorber.
 
 We modeled redshifted components from the following four bins:
 $-$0\arcsecpoint2, $-$1\arcsecpoint0, $-$1\arcsecpoint4, and $-$1\arcsecpoint8.
 The comparison of the model predictions and the observed line ratios are
 given in Tables 6 -- 9. Although the fits are generally good,
 in 3 of 4 cases the models overpredict the L$\alpha$/H$\beta$ ratio, even though
 we assumed a higher dust fraction for the red-shifted components.
 This may provide a clue to the structure of the emission-line knots.
 We assumed that the denser gas is screened by a more tenuous layer, and
 among the red-shifted components, the bulk of the L$\alpha$ arises
 in the dense gas. We have not included the effects of resonance-line
 scattering of L$\alpha$ photons by the outer, tenuous component, but
 there is enough neutral hydrogen such that the mean optical depth of
 L$\alpha$ is large ($>$ 10$^{3}$). Therefore, we believe that
 we are viewing the denser gas through a substantial column of 
 tenuous gas, in which there is additional suppression of L$\alpha$.

 An odd feature of the models is the overprediction of [Ne~III] $\lambda$3869
 in the three outer bins, since this line is generally well fit by 
 photoionization models. Furthermore, the predictions for the other
 neon lines are quite good. We do not have an explanation for this
 discrepancy.

 In addition to finding evidence for somewhat higher densities, we do not 
 find evidence that n$_{H}$ $\propto$ D$^{-2}$, as in the 
 blueshifted gas. There is a shallower fall-off ($\propto$ D$^{-1.5}$)
 between 40 and 100 pc, with the density remaining fairly
 constant out to $\sim$ 200 pc, which may be evidence of a somewhat different
 confining medium.
 
 Although our model predictions provide a good fit to the data, there
 are some open issues. First, the $-$0\arcsecpoint8 bin
 has a noticeably higher excitation spectrum than any of the
 other redshifted points (see Table 1b). This could be due to an additional
 component of ionizing radiation, due to shocks, or the absence of
 absorption between this region and the central source. Second,
 the velocities of the redshifted components are similar to those
 of the blue-shifted components at the same radial distance (CK2000).
 Hence, the force which drives these clouds outward
 does not seem to be significantly diminished by the intervening absorber. 

\subsection{SW-blueshifted Quadrant}

 Due to the weakness of the flux in the SW-blue quadrant, we
 were only able to obtain dereddened emission-line ratios for
 two bins, centered at $+$0\arcsecpoint4 and $+$1\arcsecpoint0.
 The spectra from both bins are somewhat unusual. In the $+$0\arcsecpoint4
 bin, the large He~II $\lambda$4686/H$\beta$ indicates the presence of 
 highly ionized, optically thin gas, while the [Ne~V] $\lambda$3426 and 
 [Fe~VII] $\lambda$6087 lines are much weaker than one would expect from gas 
 with an extended He$^{+2}$ zone. Also, C~IV $\lambda$1550 and 
 N~V $\lambda$1240 are quite strong, particularly compared to the weakness of 
 the [Ne~V] $\lambda$3426. The spectrum from $+$1\arcsecpoint0 bin 
 shows very strong high ionization lines, such as He~II $\lambda$4686, 
 C~IV $\lambda$1550 and N~V $\lambda$
 1240, while the [O~III] $\lambda$5007/[O~II] $\lambda$3727 ratio
 is $\approx$ 2.5, indicative of fairly low ionization gas (Ferland \&
 Netzer 1983). 

 It is possible to produce strong He~II lines at the same time as weak 
 [Ne~V] $\lambda$3426 in very highly ionized gas (U $>$ 10$^{-1}$),
 in which most of the neon is in the form of Ne~VI and higher.
 Although there will be similarly small amounts of C~IV and N~V, these
 lines will be strong relative to [Ne~V], due to their larger collision
 strengths, greater abundance (in the case of carbon), and, since these
 are resonance lines, continuum pumping. As such, we have modeled
 the $+$0\arcsecpoint4 bin as the superposition of such a highly ionized,
 optically thin component, $\approx$ 14 pc from the hidden AGN, and a low 
 ionization, optically thick component.
 Unlike our other models, the optically thick gas is not screened
 in the thin component, but is out of the plane, and ionized only
 by scattered continuum emission ($\sim$ 0.5\% of the continuum radiation
 incident on the optically thin gas). Presumably, this
 component is blocked from the AGN by an extremely optically thick
 absorber, similar to the model
 used for the hot spot (Paper II). Such a geometry
 can account for the presence of low ionization, low density gas, in
 which the [O~II] $\lambda$3727 arises, at such a small distance from the
 central source. The details and predicted line ratios are given in
 Table 10, and the fit is quite good, with the exception of [Ne~IV] 
 $\lambda$2423. Note that we have included CLOUDY90 predictions for the
 boost due to continuum pumping for L$\alpha$, N~V $\lambda$1240, 
 and C~IV $\lambda$1550, assuming a turbulent velocity of 100 km s$^{-1}$
 in the highly ionized component. Even at such low velocity, the boost is 
 significant, since $\tau$$_{line center}$ $<$ 100 for each of these lines,
 hence the pumping probability is non-negligible. 

 Due to the large uncertainties in the line fluxes,
 we have not attempted to 
 model the emission-line ratios for the bin centered at $+$1\arcsecpoint0. 
 However, the strong C~IV and N~V lines coupled with the relatively
 low ionization state indicated by the optical line ratios are
 evidence that we are again viewing a superposition
 of highly ionized gas within the emission-line bicone and 
 low ionization gas outside the bicone.

\subsection{SW-redshifted Quadrant}

 The spectra from the SW-red quadrant are characterized by a
 combination of high and low excitation emission-lines, similar to
 the NE-blue quadrant. As discussed in CK2000, 
 the kinematic profiles of the two quadrants are similar.
 Therefore, we assumed that the SW-red gas is directly ionized
 by the central continuum source. 
 
 The bins centered at $+$0\arcsecpoint0
 and $+$0\arcsecpoint2 overlap the region nearest the hot spot, and the
 models discussed in Paper II apply to these regions as well;
 therefore, we started with the bin centered at $+$1\arcsecpoint0. 
 Our predicted
 emission-line ratios fit the observed quite well, as shown in Table 12.
 We assumed n$_{H}$ $=$ 1.9 x 10$^{5}$ cm$^{-3}$ for the denser component,
 which is high enough to nearly collisionally extinguish the 
 [O~II] $\lambda$3727 line, which would explain its apparent weakness.
 Assuming n$_{H}$ $\propto$ D$^{-2}$, we also obtained a satisfactory fit for
 the bin centered at $+$1\arcsecpoint4, as shown in Table 13.
 Interestingly, this is the point where the velocity begins to 
 decrease (CK2000), although we see no effect in the emission-line ratios,
 unlike those in the NE-blue quadrant.

 The spectrum of the $+$0\arcsecpoint8 bin is somewhat different, 
 in that both the He~II $\lambda$4686/H$\beta$
 and [Ne~V] $\lambda$3426/H$\beta$ ratios are larger, which indicates a
 higher average ionization state in the gas. To model this effect, 
 we have assumed a lower density for this component than for that at
 $+$1\arcsecpoint0. 
 The results are given in Table 11, and are a reasonable fit, except for
 the UV resonance lines, although 
 there are large uncertainties for the dereddened UV line fluxes. It is
 possible that the resonance lines are enhanced by continuum pumping but,
 as noted above, one would not expect this process to be significant
 for lines that are extremely optically thick, and, in any case, this would 
 not account
 for all the observed N~V $\lambda$1240 emission. Furthermore, the
 {\it FWHM} of [O~III] $\lambda$5007 is $\sim$ 550 km s$^{-1}$ for this
 bin (CK2000), which constrains the turbulent velocity. Thus,
 although this region is apparently directly photoionized by the
 AGN, we cannot rule out other physical processes.
  
\section{Discussion}

\subsection{Physical Conditions in the inner NLR of NGC 1068}

  Based on our previous analysis and these model results, we
  can describe the conditions of the inner NLR of NGC 1068.
  The physical conditions in this region are dominated by the 
  effects of the hidden AGN, since the emission-line knots are
  photoionized by the central source and the gas is radially accelerated,
  presumably due to radiation pressure or a wind emanating from the
  nucleus (CK 2000). The small dust fractions in these clouds (generally $\sim$ 25\%)
  may indicate that they originate close to the nucleus (the
  dust sublimation radius is a few tenths of a parsec for
  a source of the luminosity of NGC 1068; Barvainis 1987). The number 
  density, n$_{H}$, of the clouds is
  generally decreasing as D$^{-2}$; as such, a given cloud should decrease
  in density by a factor of $\sim$ 100 in the time it takes to cross
  the emission-line region (e.g., from 10 pc to 100 pc). However, 
  given the thermal velocity
  within the ionized gas ($\sim$ 10$^{6}$ cm s$^{-1}$), assuming
  spherical clouds with a radius $\sim$ 10$^{16}$ cm (roughly the
  average of N$_{H}$/n$_{H}$), the density should decrease by a 
  factor of 100 in $\sim$ 1000 yrs, about 1/100 the crossing time.
  Furthermore, the high and low density components are not in pressure 
  equilibrium with each other (n$_{H}$T is typically 5 times higher for the 
  denser component).
  Hence there is likely some confinement, perhaps via
  an intercloud medium. However, the intercloud
  medium in the inner $\sim$ 100 pc must be quite tenuous, since there 
  is no evidence for cloud deceleration and the ionizing
  radiation is unattenuated except near the plane of the host galaxy.
  At $\sim$ 150 pc, the cloud velocities begin to decrease. The deceleration
  in the NE-blue quadrant is accompanied by enhanced high-ionization 
  line-emission from the clouds, which we suggest is the
  result of UV -- X-ray radiation from shocks. This supports our 
  hypothesis that the deceleration is caused by the interaction of the
  outflowing gas and the ambient medium, which is apparently
  denser here than closer to the hidden AGN.

  Regarding the cloud/medium interaction, medium resolution
  longslit STIS observations provide strong evidence for the 
  disruption of the emission-line gas, which is also attributed to shocks 
  (Cecil et al. 2000). Although the position and extent of the
  shock front is not constrained in our simple models, it is most
  likely that the shocks occur at the leading edges (the side 
  opposite the ionized face) of the outflowing emission-line clouds. 
  In any case, there is no need for additional ionizing radiation arising
  from the interaction of the radio jet and the interstellar medium, at least
  along PA 202\deg. It may still be that the jet/cloud interaction is 
  important locally; previously, we found evidence for 
  cosmic ray heating in emission-line gas which is in the 
  direction of the radio axis (Kraemer et al. 1998). 
  Unfortunately, these data provide no definitive evidence as to the nature 
  of the decelerating medium (although there is some evidence from
  the extended continuum; see below).

  We have found evidence that the ionizing continuum is heavily
  absorbed in the direction of the NE-red quadrant,
  similar to the situation
  in NGC 4151 (Alexander et al. 1999; Kraemer et al. 2000). In the
  case of NGC 4151 and several other Seyferts, Kraemer et al. (1999)
  argued that there may be absorbers between the NLR and the central source
  similar to the X-ray (Reynolds 1997; 
  George et al.
  1998) and UV (Crenshaw et al. 1999) absorbers
  often detected along the line-of-sight to their nuclei.
  However, our observations of NGC 1068 provide new insights as to the
  nature of the intervening absorber. 
  First, although the overall NLR appears to indicate
  the presence of an absorber (Alexander at al. 2000), we find that 
  the covering factor of the absorber is probably less than 0.5, since 
  much of the
  inner NLR is irradiated by an unabsorbed continuum. Second,
  the scatterer, which has a covering factor close to unity, is too highly ionized to modify the 
  SED significantly (Miller et al. 1991; Paper II). 
  Finally, it is significant that the NE-red and SW-blue quadrants,
  which show evidence of a weaker flux of ionizing radiation, lie
  close to the plane of the host galaxy (CK2000). It is not
  too surprising to find that there is more intervening material in the
  galactic plane. Hence, the absorption may be partially due to the 
  alignment of the emission-line
  bicone and the host galaxy.

\subsection{The Extended Continuum}

  In Paper I, we discussed the nature of the extended continuum emission in the
  inner NLR of NGC 1068, which we were able to deconvolve into
  two components: 1) electron-scattered light from the hidden AGN, and 2)
  an old stellar population. In addition to the region nearest the hot spot,
  there were local peaks of non-stellar continuum 
  roughly 1\arcsecpoint5 NE and SW of the hot spot. One of the slit
  positions in Axon et al.'s (1998) FOC f/48 observations, ``POS3'',
  intersects our long slit placement at approximately 1\arcsecpoint8 NE
  of the optical hot spot, where they reported the presence of excess continuum
  radiation, which they suggested was the result of shocks. However, 
  we determined that the excess UV continuum 
  had the same features as the continuum from the hot spot, 
  such as broad Fe II emission and Balmer continuum (the ``little blue bump''), and thus was
  principally scattered light. On the other hand, we have found 
  evidence for contribution of shock-induced ionizing radiation
  in this region. Are these results consistent?

  For electron scattering by an optically thin plasma, the fraction of light
  from the nucleus that is scattered by a given region is the product of the electron column density 
  of the scattering
  medium, the Thomson cross-section, and the covering factor (see discussion in Pier et al. 1994; 
  and Paper II). As shown
  in Paper I, at 2500 \AA~, the continuum at $-$1\arcsecpoint5 NE 
  is $\sim$ 1/12 the strength of that at the hotspot, and, therefore,
  1 x 10$^{-4}$ that of the hidden continuum. The maximum covering
  factor for the region subtended by the slit is $\sim$ 0.015, assuming
  a bicone opening angle of $\sim$ 80\deg~ (CK2000). If we
  impose the constraint that this gas contributes a small ($\leq$ 15\%)
  fraction of the observed H$\beta$ flux, and the gas is in photoionization
  equilibrium, the electron column 
  density is $\leq$ few x 10$^{21}$ cm$^{-2}$, which would account 
  for $\sim$ 1/3 of the continuum emission. Fast shock models predict some 
  contribution to the 
  UV continuum longward of 1200 \AA~ (Morse et al. 1996). Given the
  contraints on the total X-ray emission from the shock front discussed
  in Section 5.2, the shock could contribute $\sim$ 1/3 of the observed
  continuum at 2500 \AA~, although this would require that a significant
  fraction of the soft X-ray emission arises within a single bin along our 
  slit position, which may be unlikely. Hence, it is most probable that the 
  continuum
  emission at $-$1\arcsecpoint5 is reflected, particularly
  given the presence of the little blue bump. Although this
  requires a greater electron column density, hence denser gas, it may
  be that the scatterer is in a somewhat higher state of ionization than would 
  be predicted by simple photoionization models and, thus, contributes
  less H$\beta$ flux.

  It has been suggested that the scatterer is an X-ray absorber
  viewed across our line-of-sight (Krolik \& Kriss 1995), and certainly
  the scattering medium near the hotspot resembles such an
  absorber (Paper II). However, the knots of scattered emission further
  out (Paper I) indicate the presence of a substantial column of
  very highly ionized (U $\sim$ 1.0), high covering factor gas at 
  distances $>$ 100 pc in NGC 1068 which suggests that X-ray absorbers
  extend deep into the NLR of Seyfert galaxies, and certainly lends
  credence to the arguments for multi-zoned absorbers (Otani et al. 1996;
  Guainazzi et al. 1996; Reynolds et al. 1997). 

  Finally, there are reasons to believe the scatterer is responsible
  for the cloud deceleration\footnote[6]{In order for it to be
  as highly ionized as the scatterer, the precursor density must be
  $\sim$ a factor of 10 lower than we assumed, although its
  density was chosen primarily for illustrative purposes.}
  First the local peak in the scattered continuum 
  is coincident with the velocity turn-over (see Paper I and CK2000).
  The scattered continuum is evidence for the presence of gas to which the 
  clouds can transfer kinetic energy.  Finally, the scatterer is tenuous and 
  spatially extended, which may account 
  for the gradual deceleration of the clouds (CK2000). However, if the 
  scatterer is indeed the decelerating
  medium, it must have a lower radial velocity than the emission-line
  clouds. Also, one must explain the local build-up of such a 
  highly ionized plasma.

\section{Summary}

  We have analyzed STIS UV and optical spectra of the NLR of 
  NGC 1068. We have generated photoionization models of the NLR 
  gas and have been able to match most of the observed dereddened
  emission-line ratios. In Figure 5, we present a simplified schematic
  diagram of the NLR. As discussed in CK2000, the emission-line gas is 
  distributed in a bicone, centered on the hidden active nucleus. The bicone
  has an opening angle of $\sim$ 80\deg, and is tilted
  $\sim$ 5\deg with respect to the plane of the sky. The plane of the
  host galaxy is inclined $\sim$ 40\deg, with a major axis PA $\sim$ 106\deg.
  We have concluded the following regarding
  physical conditions in the inner $\sim$ 200 pc of the NLR, along
  PA 202\deg.

  1) The emission-lines, both NE and SW of the hot spot, show evidence of 
  two distinct kinematic components, one blueshifted with respect
  to the systemic velocity and the other redshifted, which are
  outflowing from the nucleus. The gas in
  each of these four quadrants is photoionized, for the most part,
  by continuum radiation emitted by the hidden AGN. The
  emission-line knots consist of two components of matter-bounded gas, of
  different densities;
  the denser gas may be screened in the more tenous component.
  The densities decrease with radius, but more slowly than
  expected from thermal expansion, so it is likely that there
  is some source of cloud confinement.
  There is evidence for dust mixed in with the emission-line gas,
  but the strengths of the UV resonance lines and the presence
  emission lines of magnesium and iron indicate that the
  dust/gas ratio is substantially less than in the interstellar
  medium of our Galaxy. We find no strong evidence for
  non-solar element abundances.

  2) The emission-line gas in the NE-blue quadrant appears to decelerate
  at $\sim$ 130 pc from the nucleus. At the same position, we find
  evidence for an additional source of ionization, which may be the
  UV -- X-ray radiation generated by fast ($\sim$ 1000 km s$^{-1}$)
  shocks. Although we cannot rule out such an effect in the SW-red quadrant,
  the emission-line fluxes at the point of deceleration 
  can be explained by photoionization of the gas by the AGN.

  3) The ionizing
  radiation emitted toward the NE-red quadrant is absorbed by a 
  layer of gas close to the nucleus. This absorption may be associated
  with the intersection of the emission-line bicone and the
  plane of the host galaxy which may be an important clue as to the 
  origin of high column absorbers.
  Interestingly, the radial velocities in this quadrant are similar to
  those which have an unabsorbed line-of-sight to the nucleus, which
  indicates that the source of the acceleration is not affected by 
  the absorber.

\acknowledgments

  S.B.K would like to thank John Raymond for sharing the results of 
  his shock models, and Gerald Cecil for discussing the preliminary results
  from his STIS observations of NGC 1068. We thank Cherie Miskey for help
  with the figures.
  S.B.K. and D.M.C. acknowledge support from NASA grant NAG 5-4103.

\clearpage

\figcaption[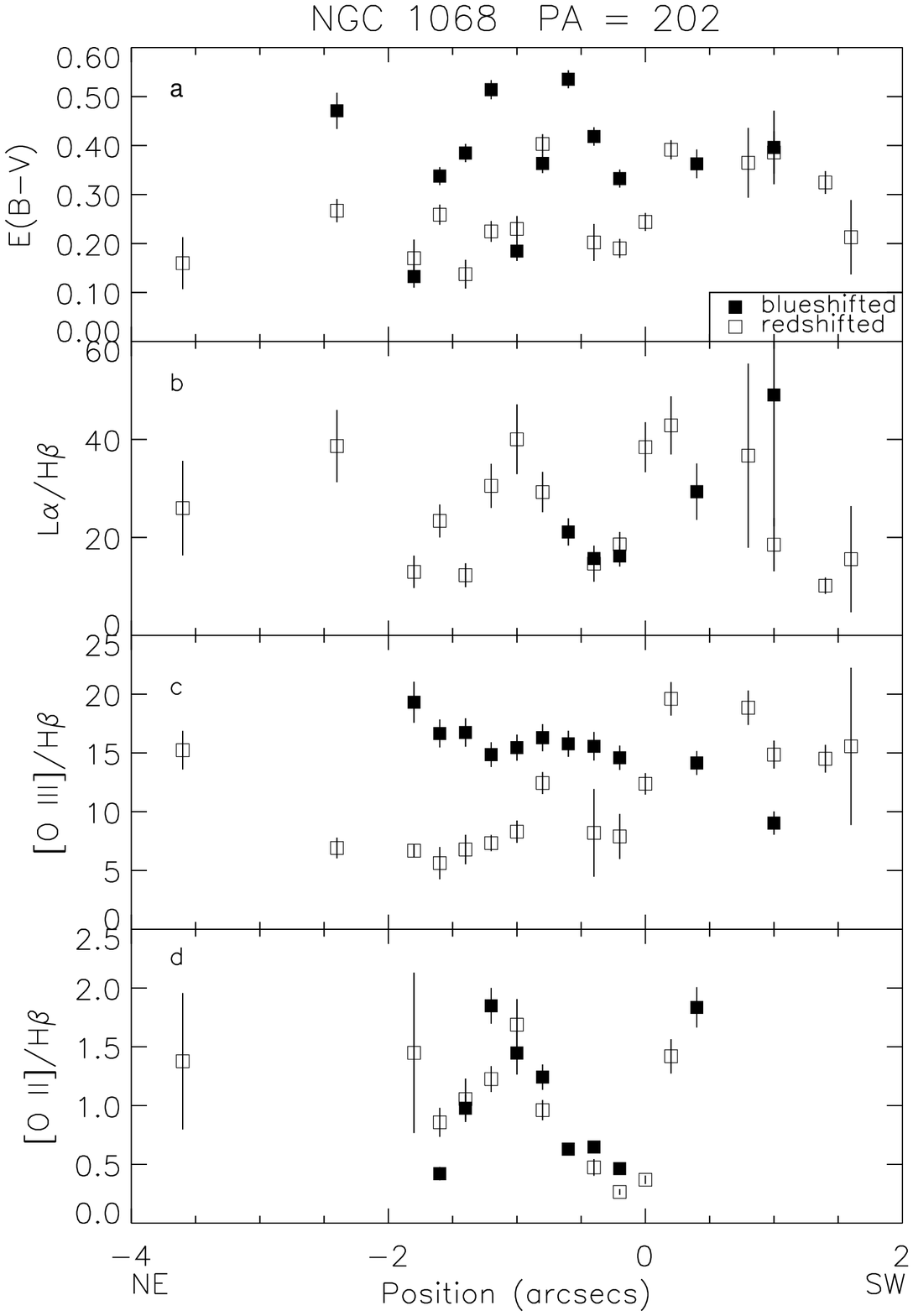]{Listed from the top to the bottom: a) the reddening 
derived from the He~II lines (see text); b)  
the dereddened L$\alpha$/H$\beta$ ratio; c) the dereddened
[O~III] $\lambda$5007/H$\beta$ ratio; d) the dereddened
[O~II] $\lambda$3727/H$\beta$ ratio (all as functions of angular distance
from the hot spot). The filled and open squares represent blueshifted and
redshifted emission, respectively (relative to systemic).
}\label{fig1} 

\figcaption[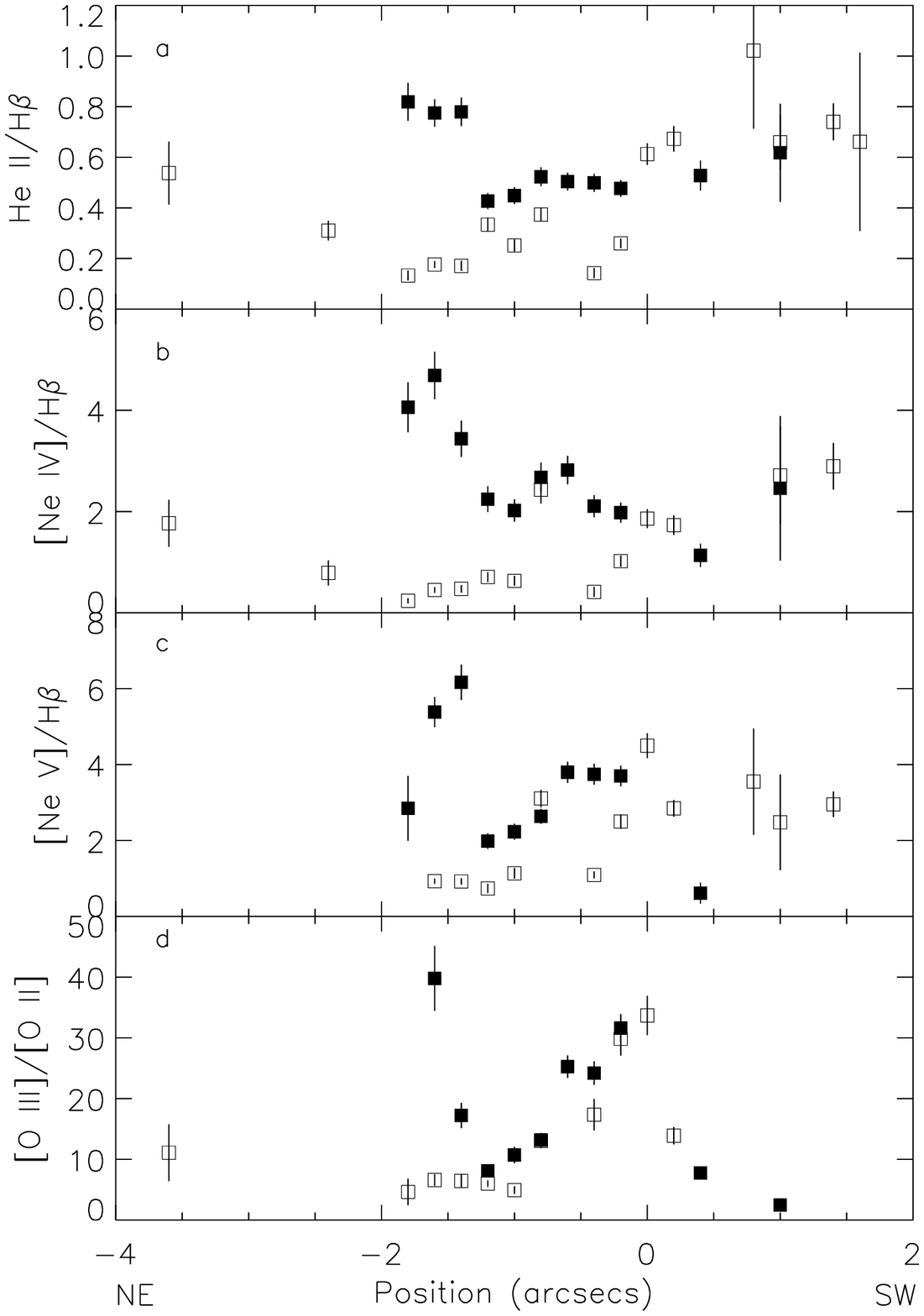]{Listed from the top to the bottom: a) 
the dereddened He~II $\lambda$4686/H$\beta$ ratio; b)  
the dereddened [Ne~IV] $\lambda$2423/H$\beta$ ratio; c) the dereddened
[Ne~V] $\lambda$3426/H$\beta$ ratio; d) the dereddened 
[O~III] $\lambda$5007/[O~II] $\lambda$3727 ratio (all as functions of 
angular distance
from the hot spot). The filled and open squares represent blueshifted and
redshifted emission, respectively (relative to systemic).
}\label{fig2} 

\figcaption[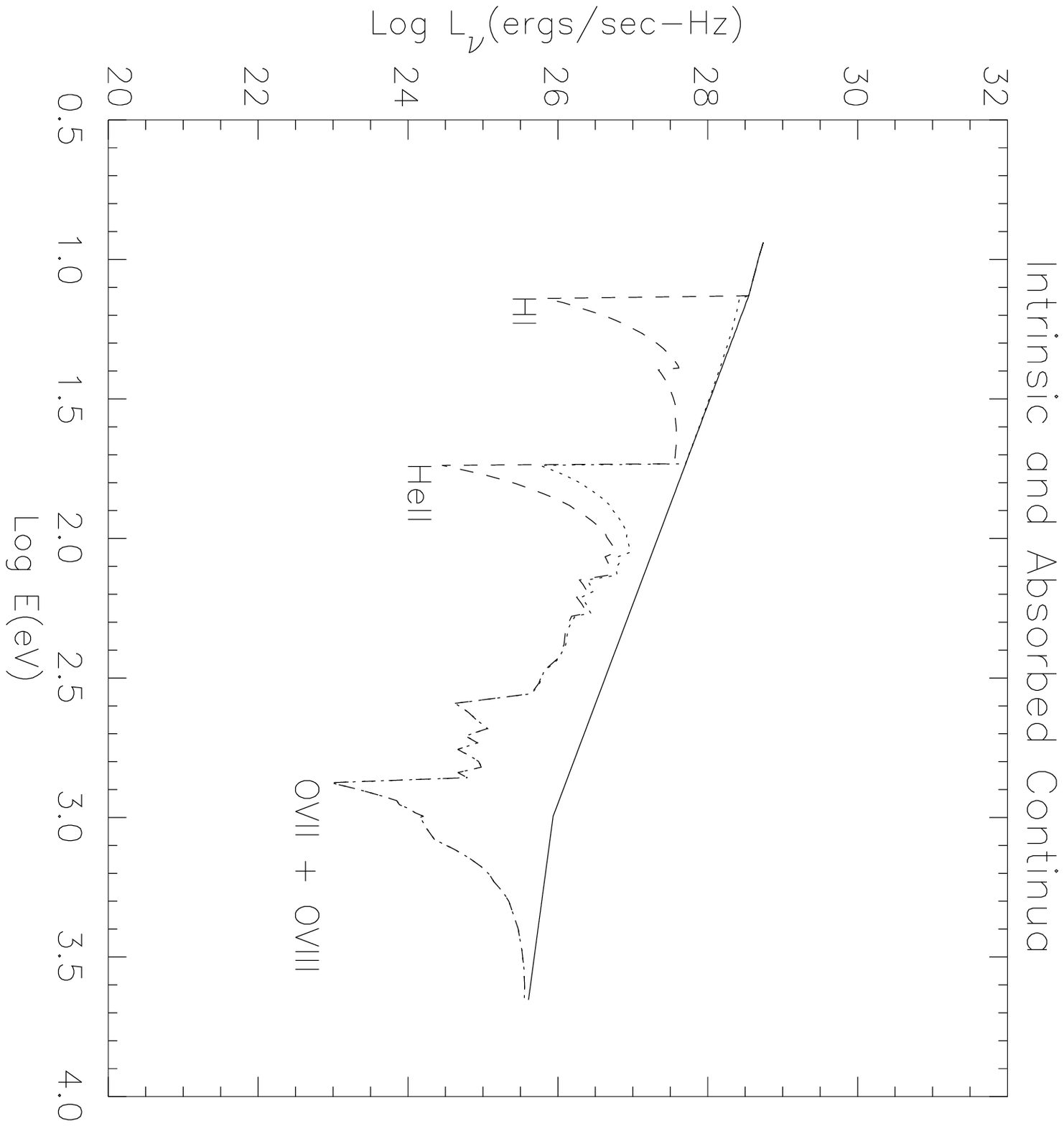]{The ionizing continua used for the non-shock models.
The solid line is the intrinsic continuum used for SW-red and
inner NE-blue quadrants. The dotted line is the continuum transmitted
though an X-ray absorber (U $=$ 1.0, N$_{H}$ $=$ 3.7 x 10$^{22}$ cm$^{-2}$).
The dashed line is the continuum used for the NE-red quandrant, 
showing the cummulative effects of the 
X-ray absorber and an additional UV absorber (U $=$ 10$^{-3.7}$, 
N$_{H}$ $=$ 1.0 x 10$^{19}$ cm$^{-2}$). The locations of the H~I, He~II,
O~VII and O~VIII ionization edges are indicated.
}\label{fig3} 

\figcaption[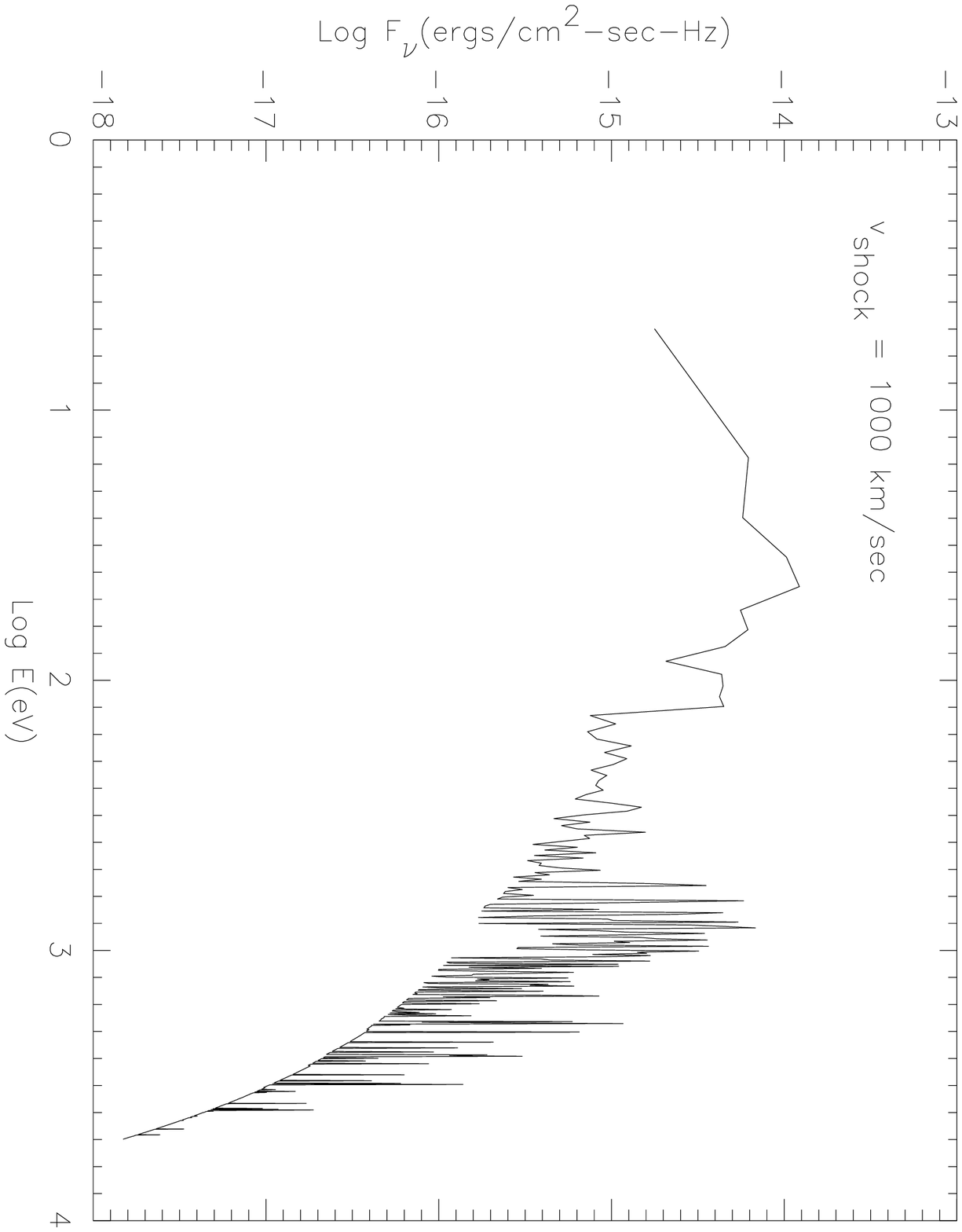]{The ionizing-radiation field produced by a 1000 km s$^{-1}$
shock. The flux scale is normalized to a preshock density of 10$^{3}$ cm$^{-3}$.
The radiation consists numerous emission lines with an underlying thermal 
bremsstrahlung continuum. The bins for energies $<$ 150 eV are too broad
to show the detailed emission-line structure, as noted in the text. (The shock model prediction
are courtesy of J.C. Raymond).
}\label{fig4} 

\figcaption[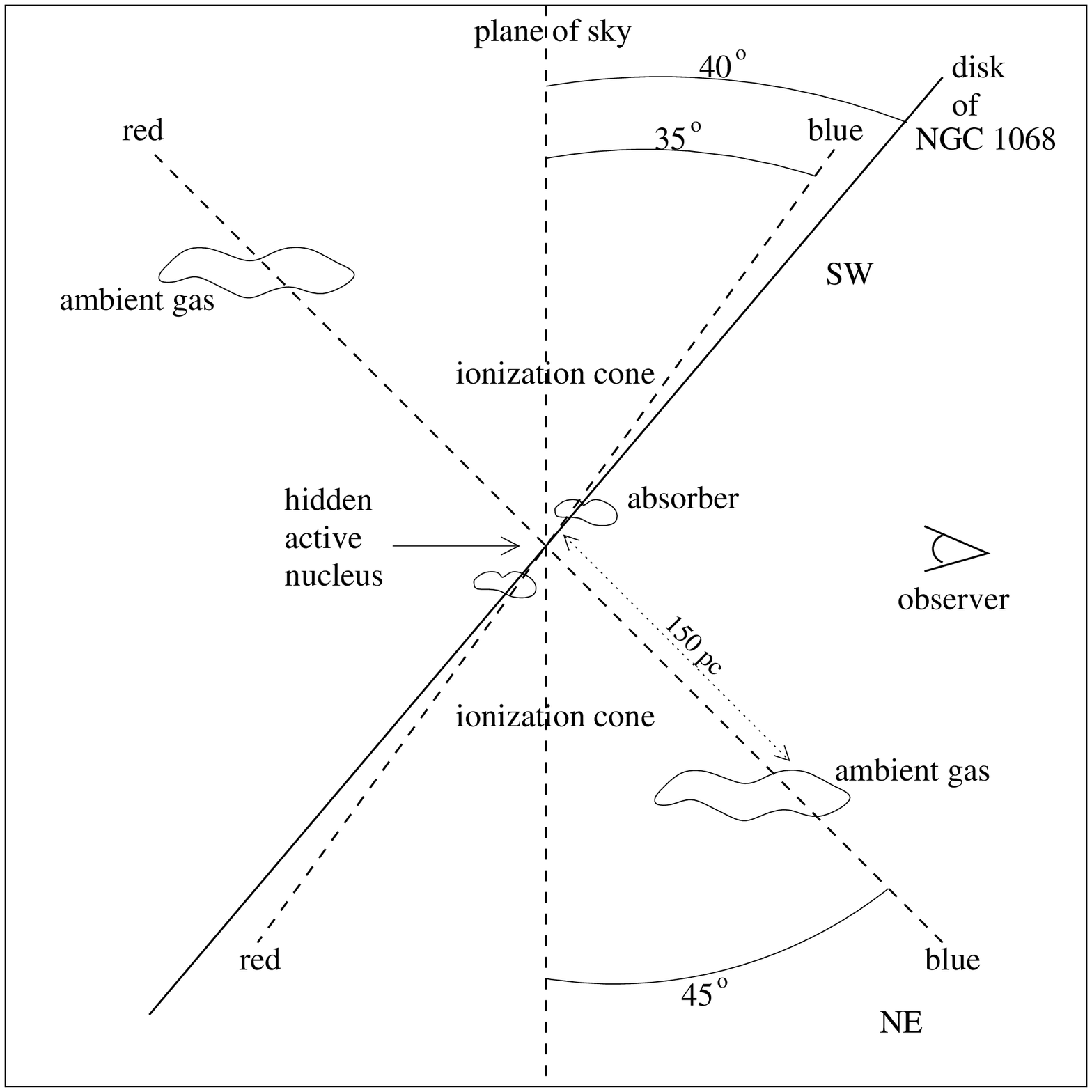]{A schematic diagram of the inner NLR of NGC 1068. 
The vertical dashed line represents the plane of the sky, the
crossed dashed lines demarcate the ionization cones, and the solid line
shows the disk of the galaxy (see CK 2000). The positions of the
blueshifted and redshifted quadrants are noted, as well as the 
general locations of the gas responsible for the continuum absorption
and the ambient gas which may slow the emission-line clouds 
(see text).  
}\label{fig5}

\clearpage
\begin{deluxetable}{lllll}
\tablenum{2}
\tablecolumns{5}
\footnotesize
\tablecaption{Line Ratios (relative to H$\beta$) from Model Components, Composite, and
Observations for $-$0\arcsecpoint2 NE, Blueshifted (D $=$ 45 pc) \label{tbl-2}}
\tablewidth{0pt}
\tablehead{
\colhead{} & \colhead{tenuous$^{a}$} & \colhead{dense$^{b}$} & 
\colhead{Composite$^{c}$} &
\colhead{Observed$^{d}$}
}
\startdata
L$\alpha$ $\lambda$1216 $+$ O~V]             &19.92             &29.73
              &24.83                         &16.82 $\pm$ 2.24 \\            
N V $\lambda$1240           	      	     & 1.64             & 0.00 
              & 0.82                         & 7.95 $\pm$ 1.03 \\
C IV $\lambda$1550          	      	     &16.73             & 0.34 
              & 8.54                         & 8.58 $\pm$ 0.89   \\
He II $\lambda$1640         	      	     & 7.06             & 0.38
              & 3.72                         & 3.45 $\pm$ 0.35  \\
$[$Ne IV] $\lambda$2423     	      	     & 2.10             & 0.02 
              & 1.06                         & 2.03 $\pm$ 0.20     \\
Mg II $\lambda$2800         	      	     & 0.00             & 1.57
              & 0.79                         & 1.13 $\pm$ 0.09      \\
$[$Ne V] $\lambda$3426      	      	     & 5.02             & 0.00 
              & 2.51                         & 3.71 $\pm$ 0.27    \\
$[$O II] $\lambda$3727      	      	     & 0.01             & 0.52
              & 0.27                         & 0.47 $\pm$ 0.03  \\
$[$Ne III] $\lambda$3869                     & 0.57             & 2.29
              & 1.43                         & 2.03 $\pm$ 0.14  \\
He II $\lambda$4686          	      	     & 0.99             & 0.06 
              & 0.52                         & 0.48 $\pm$ 0.03    \\
$[$O III] $\lambda$5007                      &13.37              &19.18 
              &16.28                         &14.66 $\pm$ 1.03    \\
$[$Fe VII] $\lambda$6087     	      	     & 0.56              & 0.00 
              & 0.28                         & 0.30 $\pm$ 0.02      \\
$[$O I] $\lambda$6300                        & 0.00              & 0.88
              & 0.44                         & 0.44 $\pm$ 0.03    \\
& & & & \\

\tablenotetext{a}{U $=$ 10$^{-1.6}$, n$_{H}$=2x10$^{4}$ cm$^{-3}$,
N$_{H}$ = 6.7 x 10$^{20}$ cm$^{-2}$; Flux$_{H\beta}$ = 1.14 ergs cm$^{-2}$ 
s$^{-1}$, dust fraction $=$ 10\%,
emitting area $=$ 1.1 x 10$^{39}$ cm$^{2}$, Depth $\geq$ 16.2 pc.}
\tablenotetext{b}{U $=$ 10$^{-2.5}$, N$_{H}$=1x10$^{5}$ cm$^{-3}$
N$_{H}$ = 1 x 10$^{21}$ cm$^{-2}$; Flux$_{H\beta}$ = 4.21 
ergs cm$^{-2}$ s$^{-1}$, dust fraction $=$ 10\%,
emitting area $=$ 3.3 x 10$^{37}$ cm$^{2}$, Depth $\geq$ 0.48 pc}
\tablenotetext{c}{50\% tenuous, 50\% dense.}
\tablenotetext{d}{dereddened; E$_{B-V}$ $=$ 0.33 $\pm$ 0.02.}

\enddata
\end{deluxetable}

\clearpage
\begin{deluxetable}{lllll}
\tablenum{3}
\tablecolumns{5}
\footnotesize
\tablecaption{Line Ratios (relative to H$\beta$) from Model Components, Composite, and
Observations for $-$1\arcsecpoint0 NE, Blueshifted (D $=$ 100 pc)\label{tbl-3}}
\tablewidth{0pt}
\tablehead{
\colhead{} & \colhead{tenuous$^{a}$} & \colhead{dense$^{b}$} & 
\colhead{Composite$^{c}$} &
\colhead{Observed$^{d}$}
}
\startdata
L$\alpha$ $\lambda$1216 $+$ O~V]             &16.27             &24.56
              &20.42                         & --  \\            
N V $\lambda$1240           	      	     & 1.08             & 0.00 
              & 0.54                         & 2.41 $\pm$ 0.34 \\
C IV $\lambda$1550          	      	     &12.50             & 0.23
              & 6.37                         & 3.48 $\pm$ 0.38   \\
He II $\lambda$1640         	      	     & 6.98             & 0.34
              & 3.66                         & 3.24 $\pm$ 0.36  \\
$[$Ne IV] $\lambda$2423     	      	     & 2.05             & 0.03 
              & 1.04                         & 2.05 $\pm$ 0.23     \\
Mg II $\lambda$2800         	      	     & 0.00             & 1.47
              & 0.74                         & 0.21 $\pm$ 0.05      \\
$[$Ne V] $\lambda$3426      	      	     & 4.12             & 0.00 
              & 2.06                         & 2.24 $\pm$ 0.21    \\
$[$O II] $\lambda$3727      	      	     & 0.02             & 1.69
              & 0.86                         & 1.45 $\pm$ 0.18  \\
$[$Ne III] $\lambda$3869                     & 0.57             & 1.99
              & 1.28                         & 1.54 $\pm$ 0.16  \\
He II $\lambda$4686          	      	     & 0.98             & 0.05
              & 0.52                         & 0.45 $\pm$ 0.03    \\
$[$O III] $\lambda$5007                      &13.15              &18.14
              &15.65                         &15.49 $\pm$ 1.11    \\
$[$Fe VII] $\lambda$6087     	      	     & 0.52              & 0.00 
              & 0.26                         & 0.28 $\pm$ 0.03      \\
$[$O I] $\lambda$6300                        & 0.00              & 1.44
              & 0.72                         & 0.70 $\pm$ 0.05    \\
& & & & \\

\tablenotetext{a}{U $=$ 10$^{-1.6}$, n$_{H}$ $=$ 4 x 10$^{3}$ cm$^{-3}$,
N$_{H}$ $=$ 6.7 x 10$^{20}$ cm$^{-2}$; Flux$_{H\beta}$ $=$ 0.24 ergs cm$^{-2}$ 
s$^{-1}$, dust fraction 10\%,
emitting area $=$ 1.89 x 10$^{38}$ cm$^{2}$, Depth $\geq$ 2.76 pc.}
\tablenotetext{b}{U $=$ 10$^{-2.5}$, n$_{H}$ $=$ 2 x 10$^{4}$ cm$^{-3}$
N$_{H}$ $=$ 3 x 10$^{21}$ cm$^{-2}$; Flux$_{H\beta}$ $=$ 0.88 
ergs cm$^{-2}$ s$^{-1}$, dust fraction 10\%,
emitting area $=$ 5.14 x 10$^{37}$ cm$^{2}$, Depth $\geq$ 0.76 pc.}
\tablenotetext{c}{50\% tenuous, 50\% dense.}
\tablenotetext{d}{dereddened; E$_{B-V}$ $=$ 0.18 $\pm$ 0.02.}

\enddata
\end{deluxetable}

\clearpage
\begin{deluxetable}{lllll}
\tablenum{4}
\tablecolumns{5}
\footnotesize
\tablecaption{Line Ratios (relative to H$\beta$) from Model Components, Composite, and
Observations for $-$1\arcsecpoint4 NE, Blueshifted (D $=$ 130 pc)
\label{tbl-4}}
\tablewidth{0pt}
\tablehead{
\colhead{} & \colhead{tenuous$^{a}$} & \colhead{dense$^{b}$} & 
\colhead{Composite$^{c}$} &
\colhead{Observed$^{d}$}
}
\startdata
L$\alpha$ $\lambda$1216 $+$ O~V]             &12.91             &41.50
              &17.20                         & --  \\            
N V $\lambda$1240           	      	     & 1.42             & 0.00 
              & 1.21                         & 5.00 $\pm$ 0.66 \\
C IV $\lambda$1550          	      	     & 5.01             & 0.26
              & 4.30                         & 5.28 $\pm$ 0.56   \\
He II $\lambda$1640         	      	     & 5.79             & 0.18
              & 4.95                         & 5.65 $\pm$ 0.58  \\
$[$Ne IV] $\lambda$2423     	      	     & 1.76             & 0.01 
              & 1.50                         & 3.53 $\pm$ 0.23     \\
Mg II $\lambda$2800         	      	     & 0.00             & 3.13
              & 0.47                         & 0.21 $\pm$ 0.05      \\
$[$Ne V] $\lambda$3426      	      	     & 6.70             & 0.00 
              & 5.70                         & 6.19 $\pm$ 0.47    \\
$[$O II] $\lambda$3727      	      	     & 0.04             & 5.21
              & 0.82                         & 0.98 $\pm$ 0.12  \\
$[$Ne III] $\lambda$3869                     & 1.23             & 3.87 
              & 1.63                         & 2.09 $\pm$ 0.17  \\
He II $\lambda$4686          	      	     & 0.88             & 0.03
              & 0.75                         & 0.78 $\pm$ 0.06    \\
$[$O III] $\lambda$5007                      &17.26              &18.54 
              &17.45                         &16.84 $\pm$ 1.22    \\
$[$Fe VII] $\lambda$6087     	      	     & 0.18              & 0.00 
              & 0.15                         & 0.40 $\pm$ 0.03      \\
$[$O I] $\lambda$6300                        & 0.00              & 3.01
              & 0.45                         & 0.30 $\pm$ 0.02    \\
& & & & \\

\tablenotetext{a}{U $=$ 10$^{-1.3}$, n$_{H}$ $=$ 3 x 10$^{3}$ cm$^{-3}$,
N$_{H}$ $=$ 2.9 x 10$^{21}$ cm$^{-2}$; Flux$_{H\beta}$ $=$ 0.55 ergs cm$^{-2}$ 
s$^{-1}$, dust fraction 25\%,
emitting area $=$ 4.45 x 10$^{38}$ cm$^{2}$, Depth $\geq$ 6.56 pc.}
\tablenotetext{b}{U $=$ 10$^{-2.9}$, n$_{H}$ $=$ 2 x 10$^{4}$ cm$^{-3}$
N$_{H}$ $=$ 6.7 x 10$^{20}$ cm$^{-2}$; Flux$_{H\beta}$ $=$ 0.29 
ergs cm$^{-2}$ s$^{-1}$, dust fraction 25\%,
emitting area $=$ 1.49 x 10$^{38}$ cm$^{2}$, Depth $\geq$ 2.19 pc.}
\tablenotetext{c}{85\% tenuous, 15\% dense.}
\tablenotetext{d}{dereddened; E$_{B-V}$ $=$ 0.38 $\pm$ 0.02.}

\enddata
\end{deluxetable}

\clearpage
\begin{deluxetable}{lllll}
\tablenum{5}
\tablecolumns{5}
\footnotesize
\tablecaption{Line Ratios (relative to H$\beta$) from Model Components, Composite, and
Observations for $-$1\arcsecpoint8 NE, Blueshifted (D $=$ 160 pc)
\label{tbl-5}}
\tablewidth{0pt}
\tablehead{
\colhead{} & \colhead{tenuous$^{a}$} & \colhead{dense$^{b}$} & 
\colhead{Composite$^{c}$} &
\colhead{Observed$^{d}$}
}
\startdata
L$\alpha$ $\lambda$1216 $+$ O~V]             &10.97             &36.86
              &13.56                         & --  \\            
N V $\lambda$1240           	      	     & 1.32             & 0.00 
              & 1.19                         & 4.53 $\pm$ 0.71 \\
C IV $\lambda$1550          	      	     & 7.13             & 0.19
              & 6.44                         & 7.94 $\pm$ 0.97   \\
He II $\lambda$1640         	      	     & 6.09             & 0.21
              & 5.50                         & 5.91 $\pm$ 0.71  \\
$[$Ne IV] $\lambda$2423     	      	     & 2.21             & 0.01
              & 1.99                         & 4.10 $\pm$ 0.50     \\
Mg II $\lambda$2800         	      	     & 0.00             & 2.64
              & 0.28                         & 0.35 $\pm$ 0.08      \\
$[$Ne V] $\lambda$3426      	      	     & 5.76             & 0.00 
              & 5.18                         & 2.85 $\pm$ 0.86    \\
$[$O II] $\lambda$3727      	      	     & 0.05             & 4.21
              & 0.47                         & --  \\
$[$Ne III] $\lambda$3869                     & 1.18             & 3.26 
              & 1.39                         & 1.61 $\pm$ 0.20  \\
He II $\lambda$4686          	      	     & 0.88             & 0.03
              & 0.79                         & 0.82 $\pm$ 0.08    \\
$[$O III] $\lambda$5007                      &18.64              &18.21 
              &18.60                         &19.35 $\pm$ 1.75    \\
$[$Fe VII] $\lambda$6087     	      	     & 0.33              & 0.00 
              & 0.29                         & 0.79 $\pm$ 0.12      \\
$[$O I] $\lambda$6300                        & 0.00              & 1.81
              & 0.18                         & 0.26 $\pm$ 0.03    \\
& & & & \\

\tablenotetext{a}{U $=$ 10$^{-1.6}$, n$_{H}$ $=$ 3 x 10$^{3}$ cm$^{-3}$,
N$_{H}$ $=$ 1.5 x 10$^{21}$ cm$^{-2}$; Flux$_{H\beta}$ $=$ 0.34 ergs cm$^{-2}$ 
s$^{-1}$, dust fraction 25\%,
emitting area $=$ 6.38 x 10$^{37}$ cm$^{2}$, Depth $\geq$ 1.0 pc.}
\tablenotetext{b}{U $=$ 10$^{-2.9}$, n$_{H}$ $=$ 2 x 10$^{4}$ cm$^{-3}$
N$_{H}$ $=$ 4.4 x 10$^{20}$ cm$^{-2}$; Flux$_{H\beta}$ $=$ 0.29 
ergs cm$^{-2}$ s$^{-1}$, dust fraction 25\%,
emitting area $=$ 8.9 x 10$^{36}$ cm$^{2}$, Depth $\geq$ 0.13 pc.}
\tablenotetext{c}{90\% tenuous, 10\% dense.}
\tablenotetext{d}{dereddened; E$_{B-V}$ $=$ 0.13 $\pm$ 0.02.}

\enddata
\end{deluxetable}

\clearpage
\begin{deluxetable}{lllll}
\tablenum{6}
\tablecolumns{5}
\footnotesize
\tablecaption{Line Ratios (relative to H$\beta$) from Model Components, Composite, and
Observations for $-$0\arcsecpoint2 NE, Redshifted (D $=$ 45 pc)
\label{tbl-6}}
\tablewidth{0pt}
\tablehead{
\colhead{} & \colhead{tenuous$^{a}$} & \colhead{dense$^{b}$} & 
\colhead{Composite$^{c}$} &
\colhead{Observed$^{d}$}
}
\startdata
L$\alpha$ $\lambda$1216 $+$ O~V]             &18.67             &45.86
              &39.06                         &18.97 $\pm$ 2.63  \\            
N V $\lambda$1240           	      	     & 2.44             & 0.00 
              & 0.61                         &10.12 $\pm$ 1.36 \\
C IV $\lambda$1550          	      	     &31.95             & 0.02
              & 8.00                         & 7.45 $\pm$ 0.79   \\
He II $\lambda$1640         	      	     & 6.64             & 0.22
              & 1.83                         & 1.88 $\pm$ 0.20  \\
$[$Ne IV] $\lambda$2423     	      	     & 2.64             & 0.00
              & 0.66                         & 1.03 $\pm$ 0.11     \\
Mg II $\lambda$2800         	      	     & 0.00             & 3.12
              & 2.34                         & 1.53 $\pm$ 0.13      \\
$[$Ne V] $\lambda$3426      	      	     & 5.97             & 0.00 
              & 1.49                         & 2.51 $\pm$ 0.19    \\
$[$O II] $\lambda$3727      	      	     & 0.03             & 0.66
              & 0.50                         & 0.27 $\pm$ 0.02  \\
$[$Ne III] $\lambda$3869                     & 0.62             & 2.80
              & 2.26                         & 2.06 $\pm$ 0.15  \\
He II $\lambda$4686          	      	     & 0.92             & 0.03
              & 0.25                         & 0.26 $\pm$ 0.02    \\
$[$O III] $\lambda$5007                      &22.22              & 4.90 
              & 9.23                         & 7.91 $\pm$ 1.91    \\
$[$Fe VII] $\lambda$6087     	      	     & 0.57              & 0.00 
              & 0.14                         & 0.39 $\pm$ 0.03      \\
$[$O I] $\lambda$6300                        & 0.00              & 0.36
              & 0.27                         & 0.30 $\pm$ 0.02    \\
& & & & \\

\tablenotetext{a}{U $=$ 10$^{-1.7}$, n$_{H}$ $=$ 4 x 10$^{3}$ cm$^{-3}$,
N$_{H}$ $=$ 3.0 x 10$^{20}$ cm$^{-2}$; Flux$_{H\beta}$ $=$ 9.65 x 10$^{-2}$ ergs cm$^{-2}$ 
s$^{-1}$, dust fraction 25\%,
emitting area $=$ 5.04 x 10$^{38}$ cm$^{2}$, Depth $\geq$ 7.42 pc.}
\tablenotetext{b}{U $=$ 10$^{-3.6}$, n$_{H}$ $=$ 2.5 x 10$^{5}$ cm$^{-3}$
N$_{H}$ $=$ 2.5 x 10$^{19}$ cm$^{-2}$; Flux$_{H\beta}$ $=$ 5.96 x 10$^{-1}$ 
ergs cm$^{-2}$ s$^{-1}$, dust fraction 50\%,
emitting area $=$ 2.45 x 10$^{38}$ cm$^{2}$, Depth $\geq$ 3.61 pc.}
\tablenotetext{c}{25\% tenuous, 75\% dense.}
\tablenotetext{d}{dereddened; E$_{B-V}$ $=$ 0.19 $\pm$ 0.02.}

\enddata
\end{deluxetable}

\clearpage
\begin{deluxetable}{lllll}
\tablenum{7}
\tablecolumns{5}
\footnotesize
\tablecaption{Line Ratios (relative to H$\beta$) from Model Components, Composite, and
Observations for $-$1\arcsecpoint0 NE, Redshifted (D $=$ 100 pc)
\label{tbl-7}}
\tablewidth{0pt}
\tablehead{
\colhead{} & \colhead{tenuous$^{a}$} & \colhead{dense$^{b}$} & 
\colhead{Composite$^{c}$} &
\colhead{Observed$^{d}$}
}
\startdata
L$\alpha$ $\lambda$1216 $+$ O~V]             &27.59             &43.06
              &39.97                         &41.10 $\pm$ 7.43  \\            
N V $\lambda$1240           	      	     & 3.25             & 0.00 
              & 0.65                         & 3.13 $\pm$ 0.55 \\
C IV $\lambda$1550          	      	     &38.95             & 0.01
              & 7.80                         & 6.80 $\pm$ 0.90   \\
He II $\lambda$1640         	      	     & 6.98             & 0.49
              & 1.79                         & 1.82 $\pm$ 0.24  \\
$[$Ne IV] $\lambda$2423     	      	     & 2.60             & 0.00
              & 0.52                         & 0.64 $\pm$ 0.10     \\
Mg II $\lambda$2800         	      	     & 0.00             & 2.30
              & 1.84                         & 2.55 $\pm$ 0.27      \\
$[$Ne V] $\lambda$3426      	      	     & 5.89             & 0.00 
              & 1.18                         & 1.14 $\pm$ 0.14    \\
$[$O II] $\lambda$3727      	      	     & 0.04             & 2.32
              & 1.86                         & 1.69 $\pm$ 0.22  \\
$[$Ne III] $\lambda$3869                     & 0.43             & 2.62
              & 2.18                         & 0.90 $\pm$ 0.12  \\
He II $\lambda$4686          	      	     & 0.96             & 0.07
              & 0.25                         & 0.25 $\pm$ 0.02    \\
$[$O III] $\lambda$5007                      &17.53             & 4.91 
              & 7.43                         & 8.33 $\pm$ 0.95    \\
$[$Fe VII] $\lambda$6087     	      	     & 0.59              & 0.00 
              & 0.12                         & 0.15 $\pm$ 0.04      \\
$[$O I] $\lambda$6300                        & 0.00              & 0.52
              & 0.42                         & 0.42 $\pm$ 0.03    \\
& & & & \\

\tablenotetext{a}{U $=$ 10$^{-1.8}$, n$_{H}$ $=$ 1 x 10$^{3}$ cm$^{-3}$,
N$_{H}$ $=$ 5.6 x 10$^{19}$ cm$^{-2}$; Flux$_{H\beta}$ $=$ 4.66 x 10$^{-3}$ ergs cm$^{-2}$ 
s$^{-1}$, dust fraction 25\%,
emitting area $=$ 2.63 x 10$^{39}$ cm$^{2}$, Depth $\geq$ 38.0 pc.}
\tablenotetext{b}{U $=$ 10$^{-3.6}$, n$_{H}$ $=$ 6.2 x 10$^{4}$ cm$^{-3}$
N$_{H}$ $=$ 2.8 x 10$^{19}$ cm$^{-2}$; Flux$_{H\beta}$ $=$ 1.66 x 10$^{-1}$ 
ergs cm$^{-2}$ s$^{-1}$, dust fraction 50\%,
emitting area $=$ 2.95 x 10$^{38}$ cm$^{2}$, Depth $\geq$ 4.35 pc.}
\tablenotetext{c}{20\% tenuous, 80\% dense.}
\tablenotetext{d}{dereddened; E$_{B-V}$ $=$ 0.18 $\pm$ 0.02.}

\enddata
\end{deluxetable}

\clearpage
\begin{deluxetable}{lllll}
\tablenum{8}
\tablecolumns{5}
\footnotesize
\tablecaption{Line Ratios (relative to H$\beta$) from Model Components, Composite, and
Observations for $-$1\arcsecpoint4 NE, Redshifted (D $=$ 130 pc)
\label{tbl-8}}
\tablewidth{0pt}
\tablehead{
\colhead{} & \colhead{tenuous$^{a}$} & \colhead{dense$^{b}$} & 
\colhead{Composite$^{c}$} &
\colhead{Observed$^{d}$}
}
\startdata
L$\alpha$ $\lambda$1216 $+$ O~V]             &24.63             &42.45
              &39.78                         &12.48 $\pm$ 2.52  \\            
N V $\lambda$1240           	      	     & 1.52             & 0.00 
              & 0.23                         & 1.28 $\pm$ 0.25 \\
C IV $\lambda$1550          	      	     &30.10             & 0.00
              & 4.52                         & 2.35 $\pm$ 0.34   \\
He II $\lambda$1640         	      	     & 6.57             & 0.36
              & 1.29                         & 1.23 $\pm$ 0.17  \\
$[$Ne IV] $\lambda$2423     	      	     & 2.81             & 0.00
              & 0.42                         & 0.48 $\pm$ 0.08     \\
Mg II $\lambda$2800         	      	     & 0.00             & 2.48
              & 2.11                         & 1.92 $\pm$ 0.20      \\
$[$Ne V] $\lambda$3426      	      	     & 4.23             & 0.00 
              & 0.63                         & 0.92 $\pm$ 0.08    \\
$[$O II] $\lambda$3727      	      	     & 0.08             & 2.80
              & 2.39                         & 1.06 $\pm$ 0.18  \\
$[$Ne III] $\lambda$3869                     & 0.86             & 2.37
              & 2.14                         & 1.06 $\pm$ 0.11  \\
He II $\lambda$4686          	      	     & 0.91             & 0.05
              & 0.18                         & 0.17 $\pm$ 0.02    \\
$[$O III] $\lambda$5007                      &24.82              & 3.35 
              & 6.57                         & 6.79 $\pm$ 1.25    \\
$[$Fe VII] $\lambda$6087     	      	     & 0.64              & 0.00 
              & 0.10                         & 0.06 $\pm$ 0.01      \\
$[$O I] $\lambda$6300                        & 0.00              & 0.46
              & 0.39                         & 0.35 $\pm$ 0.03    \\
& & & & \\

\tablenotetext{a}{U $=$ 10$^{-1.9}$, n$_{H}$ $=$ 8 x 10$^{2}$ cm$^{-3}$,
N$_{H}$ $=$ 5.9 x 10$^{19}$ cm$^{-2}$; Flux$_{H\beta}$ $=$ 4.23 x 10$^{-3}$ ergs cm$^{-2}$ 
s$^{-1}$, dust fraction 25\%,
emitting area $=$ 1.69 x 10$^{39}$ cm$^{2}$, Depth $\geq$ 24.9 pc.}
\tablenotetext{b}{U $=$ 10$^{-3.7}$, n$_{H}$ $=$ 5.0 x 10$^{4}$ cm$^{-3}$
N$_{H}$ $=$ 1.4 x 10$^{19}$ cm$^{-2}$; Flux$_{H\beta}$ $=$ 6.90 x 10$^{-2}$ 
ergs cm$^{-2}$ s$^{-1}$, dust fraction 50\%,
emitting area $=$ 5.86 x 10$^{38}$ cm$^{2}$, Depth $\geq$ 8.64 pc.}
\tablenotetext{c}{15\% tenuous, 85\% dense.}
\tablenotetext{d}{dereddened; E$_{B-V}$ $=$ 0.14 $\pm$ 0.03.}

\enddata
\end{deluxetable}

\clearpage
\begin{deluxetable}{lllll}
\tablenum{9}
\tablecolumns{5}
\footnotesize
\tablecaption{Line Ratios (relative to H$\beta$) from Model Components, Composite, and
Observations for $-$1\arcsecpoint8 NE, Redshifted (D $=$ 160 pc)
\label{tbl-9}}
\tablewidth{0pt}
\tablehead{
\colhead{} & \colhead{tenuous$^{a}$} & \colhead{dense$^{b}$} & 
\colhead{Composite$^{c}$} &
\colhead{Observed$^{d}$}
}
\startdata
L$\alpha$ $\lambda$1216 $+$ O~V]             &23.18             &42.01
              &39.19                         &13.22 $\pm$ 3.44  \\            
N V $\lambda$1240           	      	     & 0.59             & 0.00 
              & 0.09                         & 0.63 $\pm$ 0.16 \\
C IV $\lambda$1550          	      	     &21.53             & 0.00
              & 3.23                         & 1.79 $\pm$ 0.32   \\
He II $\lambda$1640         	      	     & 5.94             & 0.28
              & 1.12                         & 0.96 $\pm$ 0.17  \\
$[$Ne IV] $\lambda$2423     	      	     & 2.68             & 0.00
              & 0.40                         & 0.24 $\pm$ 0.05     \\
Mg II $\lambda$2800         	      	     & 0.01             & 1.26
              & 1.07                         & 1.57 $\pm$ 0.19      \\
$[$Ne V] $\lambda$3426      	      	     & 2.53             & 0.00 
              & 0.38                         & --    \\
$[$O II] $\lambda$3727      	      	     & 0.17             & 2.92
              & 2.51                         & 1.45 $\pm$ 0.68  \\
$[$Ne III] $\lambda$3869                     & 1.56             & 2.32
              & 2.21                         & 0.73 $\pm$ 0.10  \\
He II $\lambda$4686          	      	     & 0.83             & 0.04
              & 0.16                         & 0.13 $\pm$ 0.02    \\
$[$O III] $\lambda$5007                      &32.48              & 2.47 
              & 6.97                         & 6.70 $\pm$ 0.54    \\
$[$Fe VII] $\lambda$6087     	      	     & 0.59              & 0.00 
              & 0.09                         & --      \\
$[$O I] $\lambda$6300                        & 0.00              & 0.39
              & 0.33                         & 0.22 $\pm$ 0.03    \\
& & & & \\

\tablenotetext{a}{U $=$ 10$^{-2.0}$, n$_{H}$ $=$ 8 x 10$^{2}$ cm$^{-3}$,
N$_{H}$ $=$ 5.9 x 10$^{19}$ cm$^{-2}$; Flux$_{H\beta}$ $=$ 4.47 x 10$^{-3}$ ergs cm$^{-2}$ 
s$^{-1}$, dust fraction 25\%,
emitting area $=$ 2.36 x 10$^{39}$ cm$^{2}$, Depth $\geq$ 34.7 pc.}
\tablenotetext{b}{U $=$ 10$^{-3.8}$, n$_{H}$ $=$ 4.0 x 10$^{4}$ cm$^{-3}$
N$_{H}$ $=$ 8.9 x 10$^{18}$ cm$^{-2}$; Flux$_{H\beta}$ $=$ 3.47 x 10$^{-2}$ 
ergs cm$^{-2}$ s$^{-1}$, dust fraction 75\%,
emitting area $=$ 1.72 x 10$^{39}$ cm$^{2}$, Depth $\geq$ 25.1 pc.}
\tablenotetext{c}{15\% tenuous, 85\% dense.}
\tablenotetext{d}{dereddened; E$_{B-V}$ $=$ 0.17 $\pm$ 0.04.}

\enddata
\end{deluxetable}

\clearpage
\begin{deluxetable}{lllll}
\tablenum{10}
\tablecolumns{5}
\footnotesize
\tablecaption{Line Ratios$^{a}$ (relative to H$\beta$) from Model Components, Composite, and
Observations for $+$0\arcsecpoint4 SW, Blueshifted (D $=$ 14 pc)
\label{tbl-10}}
\tablewidth{0pt}
\tablehead{
\colhead{} & \colhead{highion$^{b}$} & \colhead{lowion$^{c}$} & 
\colhead{Composite$^{d}$} &
\colhead{Observed$^{e}$}
}
\startdata
L$\alpha$ $\lambda$1216 $+$ O~V]             &41.81 (47.20)     &28.02
              &24.59 (33.77)                 &30.57 $\pm$ 6.10  \\            
N V $\lambda$1240           	      	     & 9.78 (27.11)     & 0.00 
              & 2.93 (8.13)                  & 8.38 $\pm$ 1.62 \\
C IV $\lambda$1550          	      	     & 4.85 (11.26)     & 1.79
              & 2.71 (4.63)                  & 9.23 $\pm$ 1.32   \\
He II $\lambda$1640         	      	     & 7.70             & 1.80
              & 3.57                         & 3.82 $\pm$ 0.54  \\
$[$Ne IV] $\lambda$2423     	      	     & 0.04             & 0.41
              & 0.30                         & 1.61 $\pm$ 0.24   \\
Mg II $\lambda$2800         	      	     & 0.00             & 0.67
              & 0.47                         & 0.61 $\pm$ 0.08      \\
$[$Ne V] $\lambda$3426      	      	     & 1.74             & 0.29
              & 0.73                         & 0.61 $\pm$ 0.28  \\
$[$O II] $\lambda$3727      	      	     & 0.00             & 1.78
              & 1.25                         & 1.85 $\pm$ 0.17 \\
$[$Ne III] $\lambda$3869                     & 0.00             & 1.77
              & 1.25                         & 1.33 $\pm$ 0.11  \\
He II $\lambda$4686          	      	     & 0.99             & 0.26
              & 0.48                         & 0.53 $\pm$ 0.06    \\
$[$O III] $\lambda$5007                      & 0.01              &20.76
              &14.53                         &14.22 $\pm$ 1.02    \\
$[$Fe VII] $\lambda$6087     	      	     & 0.00              & 0.13 
              & 0.09                         & 0.09 $\pm$ 0.03      \\
$[$O I] $\lambda$6300                        & 0.00              & 0.56
              & 0.39                         & 0.43 $\pm$ 0.04    \\
& & & & \\

\tablenotetext{a}{Values in parentheses are for a turbulent velocity of 100
km s$^{-1}$ (see discussion in text).}
\tablenotetext{b}{U $=$ 10$^{-0.45}$, n$_{H}$ $=$ 1.5 x 10$^{4}$ cm$^{-3}$,
N$_{H}$ $=$ 5.0 x 10$^{20}$ cm$^{-2}$; Flux$_{H\beta}$ $=$ 0.37 ergs cm$^{-2}$ 
s$^{-1}$, no dust,
emitting area $=$ 1.22 x 10$^{38}$ cm$^{2}$, Depth $\geq$ 1.80 pc.}
\tablenotetext{c}{U $=$ 10$^{-2.1}$, n$_{H}$ $=$ 3.0 x 10$^{3}$ cm$^{-3}$
N$_{H}$ $=$ 2.0 x 10$^{21}$ cm$^{-2}$; Flux$_{H\beta}$ $=$ 0.30
ergs cm$^{-2}$ s$^{-1}$, dust fraction 10\%,
emitting area $=$ 3.57 x 10$^{38}$ cm$^{2}$, Depth $\geq$ 5.25 pc.}
\tablenotetext{d} {30\% highion, 70\% lowion.}
\tablenotetext{e}{dereddened; E$_{B-V}$ $=$ 0.36 $\pm$ 0.03.}

\enddata
\end{deluxetable}

\clearpage
\begin{deluxetable}{lllll}
\tablenum{11}
\tablecolumns{5}
\footnotesize
\tablecaption{Line Ratios (relative to H$\beta$) from Model Components, Composite, and
Observations for $+$0\arcsecpoint8 SW, Redshifted (D $=$ 40 pc)
\label{tbl-11}}
\tablewidth{0pt}
\tablehead{
\colhead{} & \colhead{tenuous$^{a}$} & \colhead{dense$^{b}$} & 
\colhead{Composite$^{c}$} &
\colhead{Observed$^{d}$}
}
\startdata
L$\alpha$ $\lambda$1216 $+$ O~V]             &12.63             &28.35
              &14.99                         &38.25 $\pm$20.06  \\            
N V $\lambda$1240           	      	     & 1.10             & 0.00 
              & 0.94                         &21.52 $\pm$10.79 \\
C IV $\lambda$1550          	      	     & 9.45             & 0.10
              & 8.05                         &10.76 $\pm$ 3.67   \\
He II $\lambda$1640         	      	     & 6.21             & 0.17
              & 5.30                         & 7.40 $\pm$ 2.42  \\
$[$Ne IV] $\lambda$2423     	      	     & 3.04             & 0.00
              & 1.73                         & --     \\
Mg II $\lambda$2800         	      	     & 0.00             & 2.19
              & 0.33                         & 1.12 $\pm$ 0.52      \\
$[$Ne V] $\lambda$3426      	      	     & 4.62             & 0.00 
              & 3.93                         & 3.57 $\pm$ 1.41  \\
$[$O II] $\lambda$3727      	      	     & 0.02             & 0.42
              & 0.08                         & --  \\
$[$Ne III] $\lambda$3869                     & 1.12             & 2.41
              & 1.31                         & 2.10 $\pm$ 0.54  \\
He II $\lambda$4686          	      	     & 0.89             & 0.02
              & 0.76                         & 1.03 $\pm$ 0.31    \\
$[$O III] $\lambda$5007                      &19.10              &12.07
              &18.05                         &18.95 $\pm$ 1.46    \\
$[$Fe VII] $\lambda$6087     	      	     & 0.47              & 0.00 
              & 0.40                         & 0.24 $\pm$ 0.12      \\
$[$O I] $\lambda$6300                        & 0.00              & 1.57
              & 0.24                         & --    \\
& & & & \\

\tablenotetext{a}{U $=$ 10$^{-1.6}$, n$_{H}$ $=$ 2.3 x 10$^{4}$ cm$^{-3}$,
N$_{H}$ $=$ 8.5 x 10$^{20}$ cm$^{-2}$; Flux$_{H\beta}$ $=$ 1.64 ergs cm$^{-2}$ 
s$^{-1}$, dust fraction 25\%,
emitting area $=$ 1.48 x 10$^{37}$ cm$^{2}$, Depth $\geq$ 0.22 pc.}
\tablenotetext{b}{U $=$ 10$^{-2.9}$, n$_{H}$ $=$ 2.3 x 10$^{5}$ cm$^{-3}$
N$_{H}$ $=$ 6.4 x 10$^{20}$ cm$^{-2}$; Flux$_{H\beta}$ $=$ 3.53 
ergs cm$^{-2}$ s$^{-1}$, dust fraction 25\%,
emitting area $=$ 1.21 x 10$^{36}$ cm$^{2}$, Depth $\geq$ 0.018 pc.}
\tablenotetext{c}{85\% tenuous, 15\% dense.}
\tablenotetext{d}{dereddened; E$_{B-V}$ $=$ 0.36 $\pm$ 0.04.}

\enddata
\end{deluxetable}

\clearpage
\begin{deluxetable}{lllll}
\tablenum{12}
\tablecolumns{5}
\footnotesize
\tablecaption{Line Ratios (relative to H$\beta$) from Model Components, Composite, and
Observations for $+$1\arcsecpoint0 SW, Redshifted (D $=$ 55 pc)
\label{tbl-12}}
\tablewidth{0pt}
\tablehead{
\colhead{} & \colhead{tenuous$^{a}$} & \colhead{dense$^{b}$} & 
\colhead{Composite$^{c}$} &
\colhead{Observed$^{d}$}
}
\startdata
L$\alpha$ $\lambda$1216 $+$ O~V]             &14.90             &31.64
              &19.09                         &19.36 $\pm$ 5.80  \\            
N V $\lambda$1240           	      	     & 0.75             & 0.00 
              & 0.56                         & 4.99 $\pm$ 1.44 \\
C IV $\lambda$1550          	      	     &10.27             & 0.05
              & 7.72                         & 6.12 $\pm$ 1.27   \\
He II $\lambda$1640         	      	     & 6.61             & 0.29
              & 5.03                         & 4.77 $\pm$ 1.01  \\
$[$Ne IV] $\lambda$2423     	      	     & 2.26             & 0.00
              & 1.70                         & 2.79 $\pm$ 1.00     \\
Mg II $\lambda$2800         	      	     & 0.00             & 2.18
              & 0.55                         & 0.87 $\pm$ 0.29      \\
$[$Ne V] $\lambda$3426      	      	     & 3.74             & 0.00 
              & 2.81                         & 2.49 $\pm$ 1.27  \\
$[$O II] $\lambda$3727      	      	     & 0.02             & 0.56
              & 0.16                         & --  \\
$[$Ne III] $\lambda$3869                     & 1.02             & 2.31
              & 1.34                         & 1.78 $\pm$ 0.33  \\
He II $\lambda$4686          	      	     & 0.94             & 0.04
              & 0.72                         & 0.66 $\pm$ 0.11    \\
$[$O III] $\lambda$5007                      &17.67              &10.85
              &15.97                         &14.94 $\pm$ 1.20    \\
$[$Fe VII] $\lambda$6087     	      	     & 0.55              & 0.00 
              & 0.41                         & --      \\
$[$O I] $\lambda$6300                        & 0.00              & 1.24
              & 0.31                         & 0.21 $\pm$ 0.03    \\
& & & & \\

\tablenotetext{a}{U $=$ 10$^{-1.8}$, n$_{H}$ $=$ 1.9 x 10$^{4}$ cm$^{-3}$,
N$_{H}$ $=$ 4.7 x 10$^{20}$ cm$^{-2}$; Flux$_{H\beta}$ $=$ 0.80 ergs cm$^{-2}$ 
s$^{-1}$, dust fraction 25\%,
emitting area $=$ 5.28 x 10$^{37}$ cm$^{2}$, Depth $\geq$ 0.78 pc.}
\tablenotetext{b}{U $=$ 10$^{-3.0}$, n$_{H}$ $=$ 1.9 x 10$^{5}$ cm$^{-3}$
N$_{H}$ $=$ 4.4 x 10$^{20}$ cm$^{-2}$; Flux$_{H\beta}$ $=$ 2.43
ergs cm$^{-2}$ s$^{-1}$, dust fraction 25\%,
emitting area $=$ 5.79 x 10$^{36}$ cm$^{2}$, Depth $\geq$ 0.085 pc.}
\tablenotetext{c}{75\% tenuous, 25\% dense.}
\tablenotetext{d}{dereddened; E$_{B-V}$ $=$ 0.39 $\pm$ 0.04.}

\enddata
\end{deluxetable}

\clearpage
\begin{deluxetable}{lllll}
\tablenum{13}
\tablecolumns{5}
\footnotesize
\tablecaption{Line Ratios (relative to H$\beta$) from Model Components, Composite, and
Observations for $+$1\arcsecpoint4 SW, Redshifted (D $=$ 85 pc)
\label{tbl-13}}
\tablewidth{0pt}
\tablehead{
\colhead{} & \colhead{tenuous$^{a}$} & \colhead{dense$^{b}$} & 
\colhead{Composite$^{c}$} &
\colhead{Observed$^{d}$}
}
\startdata
L$\alpha$ $\lambda$1216 $+$ O~V]             &13.71             &29.96
              &17.77                         &10.52 $\pm$ 1.76  \\            
N V $\lambda$1240           	      	     & 0.64             & 0.00 
              & 0.48                         & 2.99 $\pm$ 0.49 \\
C IV $\lambda$1550          	      	     & 9.00             & 0.05
              & 6.76                         & 4.44 $\pm$ 0.56   \\
He II $\lambda$1640         	      	     & 6.73             & 0.32
              & 5.13                         & 5.36 $\pm$ 0.66  \\
$[$Ne IV] $\lambda$2423     	      	     & 2.27             & 0.01
              & 1.71                         & 2.96 $\pm$ 0.47     \\
Mg II $\lambda$2800         	      	     & 0.00             & 1.79
              & 0.45                         & --      \\
$[$Ne V] $\lambda$3426      	      	     & 3.50             & 0.00 
              & 2.63                         & 2.96 $\pm$ 0.34  \\
$[$O II] $\lambda$3727      	      	     & 0.03             & 0.98
              & 0.27                         & --  \\
$[$Ne III] $\lambda$3869                     & 0.88             & 2.03
              & 1.17                         & 1.28 $\pm$ 0.12  \\
He II $\lambda$4686          	      	     & 0.96             & 0.05
              & 0.73                         & 0.74 $\pm$ 0.07    \\
$[$O III] $\lambda$5007                      &15.92              &11.87
              &14.91                         &14.58 $\pm$ 1.18    \\
$[$Fe VII] $\lambda$6087     	      	     & 0.53              & 0.00 
              & 0.40                         & 0.13 $\pm$ 0.04      \\
$[$O I] $\lambda$6300                        & 0.00              & 0.41
              & 0.10                         & 0.08 $\pm$ 0.01    \\
& & & & \\

\tablenotetext{a}{U $=$ 10$^{-1.7}$, n$_{H}$ $=$ 8.1 x 10$^{3}$ cm$^{-3}$,
N$_{H}$ $=$ 4.5 x 10$^{20}$ cm$^{-2}$; Flux$_{H\beta}$ $=$ 0.33 ergs cm$^{-2}$ 
s$^{-1}$, dust fraction 25\%,
emitting area $=$ 3.44 x 10$^{38}$ cm$^{2}$, Depth $\geq$ 5.07 pc}
\tablenotetext{b}{U $=$ 10$^{-3.0}$, n$_{H}$ $=$ 8.1 x 10$^{4}$ cm$^{-3}$
N$_{H}$ $=$ 1.5 x 10$^{20}$ cm$^{-2}$; Flux$_{H\beta}$ $=$ 1.05
ergs cm$^{-2}$ s$^{-1}$, dust fraction 25\%,
emitting area $=$ 3.62 x 10$^{37}$ cm$^{2}$, Depth $\geq$ 0.53 pc.}
\tablenotetext{c}{75\% tenuous, 25\% dense.}
\tablenotetext{d}{dereddened; E$_{B-V}$ $=$ 0.32 $\pm$ 0.02.}

\enddata
\end{deluxetable}

\clearpage
\plotone{fig1.ps}

\clearpage
\plotone{fig2.ps}

\clearpage
\plotone{fig3.ps}

\clearpage
\plotone{fig4.ps}

\clearpage
\plotone{fig5.ps}


\clearpage			       
\begin{references}


\reference{ale1999}Alexander, T., Sturm, E., Lutz, D., Sternberg, A.,
Netzer, H., \& Genzel, R. 1999, \apj, 512, 204

\reference{ale2000}Alexander, T., Lutz, D., Sturm, E., Genzel, R.,
Sternberg, A., \& Netzer, H. 2000, \apj, in press

\reference{all1998}Allen, M.G., Dopita, M.A, \& Tsevtanov, Z.I. 1998, \apj, 
493, 571

\reference{axo1998}Axon, D.J., Marconi, A., Capetti, A., Macchetto, F.D.,
Schreier, E., \& Robinson, A. 1998, \apj, 496, L75

\reference{ant1985}Antonucci, R.R, \& Miller, J.S. 1985, \apj, 297, 621 

\reference{ant1993}Antonucci, R.R. 1993, ARA\&A, 31, 64 

\reference{bar1987}Barvainis, R., 1987, \apj, 320, 537

\reference{bal1997}Bland-Hawthorn, J., Gallimore, J.F., Tacconi, L.J.,
Brinks, E., Baum, S.A., Antonucci, R.R.J., \& Cecil, G.N. 1997, Ap\&SS, 248, 9

\reference{bri1997}Brinks, E., Skillman, E.D., Terlevich, R.J., \& Terlevich,
E.T. 1997, Ap\&SS, 248, 23

\reference{cap1997}Capetti, A., Macchetto, F.D. , \& Lattanzi, M.G. 
1997, Ap\&SS, 248, 127

\reference{cec2000}Cecil, G. , et al. 2000, in preparation

\reference{cre2000}Crenshaw, D.M., \& Kraemer, S.B. 2000a, \apj, in press 
(Paper I)

\reference{cre2000b}Crenshaw, D.M., \& Kraemer, S.B. 2000b, \apj,
532, L101 (CK2000)

\reference{cre1999}Crenshaw, D.M., Kraemer, S.B., Boggess, A., Maran, S.P.,
Mushotzky, R.F., \& Wu, C.-C. 1999, \apj, 561, 750

\reference{dra1984}Draine, B.T, \& Lee, H.M. 1984, \apj, 285, 89

\reference{eva1986}Evans, I.N., \& Dopita, M.A. 1986, \apj, 310, L15

\reference{eva1991}Evans, I.N., Ford, H.C., Kinney, A.L., Antonucci, R.R.J.,
Armus, L., \& Caganoff, S. 1991, \apj, 369, L27

\reference{fer1995}Ferguson, J.W., Ferland, G.J., \& Pradhan, A.K. 1991, \apj, 
438, L55 

\reference{fer1983}Ferland, G.J., \& Netzer, H. 1983, \apj, 264, 105

\reference{fer1998}Ferland, G.J., et al. 1998, PASP, 110, 761

\reference{gal1996}Gallimore, J.F., Baum, S.A., \& O'Dea, C.P. 1996, \apj, 
464, 198
\reference{gal1997}Gallimore, J.F., \& Tacconi, L.J. 1997, Ap\&SS, 248, 253

\reference{geo1998}George, I.M., Turner, T.J., Netzer, H., Nandra, K.,
Mushitzky, R.F., \& Yaqoob, T. 1998, ApJS, 114, 73

\reference{gre1989}Grevesse, N., \& Anders, E. 1989, in AIP Conf. Proc. 183,
Cosmic Abundances of Matter, ed. C.J. Waddington (New York: AIP), 1

\reference{gua1996}Guainazzi, M., Mihara, T., Otani, C., \& Matsouka, M.
1996, PASJ, 48, 781

\reference{kra1985}Kraemer, S.B. 1985, Ph.D. thesis, Univ. Maryland

\reference{kra2000}Kraemer, S.B., \& Crenshaw, D.M. 2000, \apj, in press (Paper
II)

\reference{kra2000}Kraemer, S.B., Crenshaw, D.M., Hutchings, J.B., Gull, T.R.,
Kaiser, M.E., Nelson, C.H., Weistrop, D. 2000, \apj, 531, 278

\reference{kra1986}Kraemer, S.B., \& Harrington, J.P. 1986, \apj, 307, 478

\reference{kra1998}Kraemer, S.B., Ruiz, J.R., \& Crenshaw. D.M. 1998,
\apj, 508, 232 

\reference{kra1999}Kraemer, S.B., Turner, T.J., Crenshaw, D.M, \& George,
I.M. 1999, \apj, 519, 69

\reference{kra1994}Kraemer, S.B., Wu, C.-C., Crenshaw, D.M., \& Harrington,
J.P. 1994, \apj, 435, 171

\reference{kri1992}Kriss, G.A., et al. 1992, \apj, 394, L37 

\reference{kro1995}Krolik, J.H., \& Kriss, G.A. 1995, \apj, 447, 512

\reference{mac1994}Macchetto, F., Capetti, A., Sparks, W.B., Axon, D.J.,
\& Boksenberg, A. 1994, \apj, 435, L15

\reference{mil1991}Miller, J.S., Goodrich, R.W., \& Mathews, W.G. 1991, \apj,
378, 47 

\reference{mor1996}Morse, J.A., Raymond, J.C., \& Wilson, A.S. 1996, PASP,
108, 426

\reference{net1997}Netzer, H. 1997, Ap\&SS, 248, 127 

\reference{net1997}Netzer, H., \& Turner, T.J. 1997, \apj, 488, 694 

\reference{ost1989}Osterbrock, D.E. 1989, Astrophysics of Gaseous Nebulae
and Active Galactic Nuclei (Mill Valley, Univ. Science Books)

\reference{ota1996}Otani, C., et al. 1996, PASJ, 48, 211

\reference{pie1994}Pier, E.A., et al. 1994, \apj, 428, 124 

\reference{pog1988}Pogge, R.W. 1988, \apj, 328, 519

\reference{rey1997}Reynolds, C.S. 1997, MNRAS, 286, 513

\reference{rey1997}Reynolds, C.S, Ward, M.J., Fabian, A.C., \& Celotti, A.
1997, MNRAS, 291, 403

\reference{sav1979}Savage, B.D., \& Mathis, J.S. 1979, ARA\&A, 17, 73

\reference{sch1993}Schulz, H., \& Komossa, S. 1993, A\&A, 278, 29

\reference{shu1985}Shull, J.M., \& Van Steenberg, M.E. 1985, \apj, 294, 599

\reference{sno1996}Snow, T.P., \& Witt, A.N. 1996, \apj, 468, L68

\reference{sut1993}Sutherland, R.S., Bicknell, G.V., \& Dopita, M.A. 1993, 
\apj, 414, 510

\reference{tur1997}Turner, T.J., George, I.M., Nandra, K., \& Mushotzky, R.F.
1997, \apj, 488, 164

\reference{vie1989}Viegas-Aldrovandi, S.M., \& Contini, M. 1989a, A\&A, 215, 253

\reference{vie1989}Viegas-Aldrovandi, S.M., \& Contini, M. 1989b, \apj, 
339, 689

\reference{wil1992}Wilson, A.S., Elvis, M., Lawrence, A., \& Bland-Hawthorn,
J. 1992, \apj, 391, L75

\reference{wil1999}Wilson, A.S., \& Raymond, J.C. 1999, \apj, 513, L115

\reference{wil1983}Wilson, A.S., \& Ulvestad, J.A. 1983, \apj, 275, 8

\end{references}
\end{document}